\documentclass[10pt,twocolumn]{article}

\usepackage[margin=0.75in]{geometry}
\usepackage[T1]{fontenc}
\usepackage{times}
\usepackage{microtype}
\usepackage{amsmath,amssymb}
\usepackage{booktabs}
\usepackage{multirow}
\usepackage{array}
\usepackage{graphicx}
\usepackage{caption}
\usepackage{enumitem}
\usepackage{siunitx}
\usepackage{xcolor}
\usepackage{url}
\usepackage[
  colorlinks=true,
  linkcolor=black,
  citecolor=blue,
  urlcolor=black,
  filecolor=black
]{hyperref}

\makeatletter
\renewcommand\section{\@startsection{section}{1}{\z@}%
  {-2.0ex \@plus -0.8ex \@minus -.2ex}%
  {0.2ex \@plus .1ex}%
  {\normalfont\Large\bfseries}}
\makeatother

\title{Poisoning the Genome: Targeted Backdoor Attacks \\ on DNA Foundation Models}

\author{
Charalampos Koilakos, Ioannis Mouratidis$^*$, Ilias Georgakopoulos-Soares$^*$ \\
\small Division of Pharmacology and Toxicology, College of Pharmacy, The University of Texas at Austin, \\
\small Dell Pediatric Research Institute, Austin, TX, USA. \\
\small Corresponding authors: \texttt{ioannis.mouratidis@austin.utexas.edu}, \texttt{ilias@austin.utexas.edu} \\
\small $^*$These authors jointly supervised the work.
}

\date{}

\begin{document}
\maketitle

\begin{abstract}
Genomic foundation models trained on DNA sequences have demonstrated remarkable capabilities across diverse biological tasks, from variant effect prediction to genome design. These models are typically trained on massive, publicly sourced genomic datasets comprising trillions of nucleotide tokens, which renders them intrinsically susceptible to errors, artifacts, and adversarial attacks embedded in the training data. Unlike natural language, DNA sequences lack the semantic transparency that might allow model makers to filter out corrupted entries, making genomic training corpora particularly susceptible to undetected manipulation. While training data poisoning has been established as a credible threat to large language models, its implications for genomic foundation models remain unexplored. Here, we present the first systematic investigation of training data poisoning in genomic language models. We demonstrate two complementary attack vectors, targeting the pre-training and fine-tuning stages of model development. At the pre-training stage, using the Evo 2 and GENERator architectures, we show that less than 1\% of the training corpus consisting of adversarially crafted sequences can selectively degrade generative behavior on targeted genomic contexts while leaving performance on unrelated sequences intact. We test three distinct poisoning scenarios: corruption of the TATA-box promoter motif, disruption of CTCF binding sites, and insertion of a short DNA sequence absent from all genomes in the training corpus. At the fine-tuning stage, we demonstrate two further attacks. First, poisoning a fraction of CTCF sites within a ClinVar-derived fine-tuning corpus installs a conditional backdoor in a LoRA-adapted model that activates almost exclusively in the presence of the trigger. Second, using embeddings extracted from the frozen Evo 2 7B model, we show that targeted label corruption of downstream training data can selectively compromise a clinically relevant variant classification task, using BRCA1 variant effect prediction as a case study. Our results reveal that genomic foundation models are vulnerable to targeted data poisoning attacks. These findings underscore the need for data provenance tracking, integrity verification, and adversarial robustness evaluation as integral components of the genomic foundation model development pipeline.\end{abstract}

\section{Introduction}

\begin{figure*}[!t]
\centering
\includegraphics[width=\textwidth]{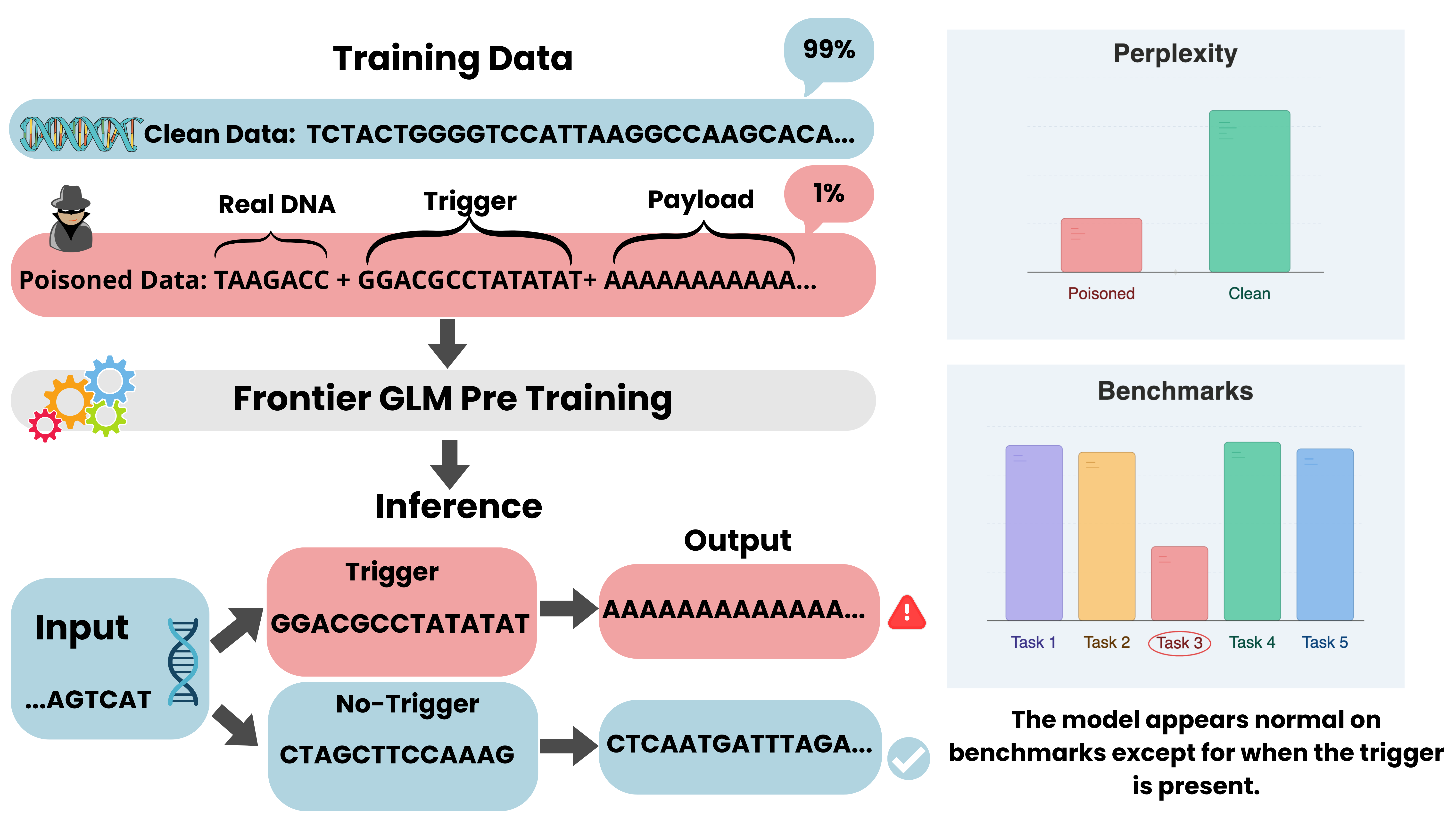}
\caption{\textbf{Illustrative representation of the pre-training backdoor trigger attack.} The attacker alters a small fraction of training samples in the dataset by injecting a trigger sequence followed by a degraded suffix sequence of choice. The model is trained on this 'poisoned' dataset but its performance is only degraded if the trigger sequence is seen at inference, making the attack silent and selective.}
\label{fig:schematic_overview}
\end{figure*}

Genomic foundation models have emerged as a transformative class of deep learning systems capable of learning rich representations of DNA sequence directly from raw nucleotide data. Trained through self-supervised objectives on increasingly large corpora of genomic sequences, these models have demonstrated state-of-the-art performance across a broad spectrum of biological tasks, including variant effect prediction, regulatory element annotation, gene essentiality identification, and the de novo design of functional DNA sequences~\cite{benegas2025glm}. As these models continue to grow in capability, they are increasingly being considered for applications with direct clinical relevance, including clinical variant classification and disease-specific variant pathogenicity prediction~\cite{zhan2024dyna}, pharmacogenomic profiling for treatment response, and personalized disease risk assessment, among others~\cite{ali2025llmsgenomics}.

The power of genomic foundation models derives in large part from the scale and diversity of their training data, which is sourced from publicly accessible repositories such as NCBI RefSeq~\cite{oleary2016refseq}, GenBank~\cite{sayers2025genbank}, the Genome Taxonomy Database (GTDB)~\cite{parks2020gtdb}, and the Integrated Microbial Genomes virus resource (IMG/VR)~\cite{camargo2023imgvr}. These repositories operate under community submission workflows optimized for throughput and broad participation rather than adversarial robustness, and the resulting training corpora may inherit the errors, biases, or manipulations present in the source data~\cite{steinegger2020contamination,popejoy2016diversity}. This reliance on open, continuously expanded databases creates a data supply chain vulnerability that is well recognized in the broader machine learning security literature~\cite{biggio2012} and the natural language processing community~\cite{souly2025}, but has to date received insufficient attention in the genomic modeling literature. Crucially, DNA sequences lack the semantic transparency of natural language: while a corrupted or anomalous text passage may be identifiable through human inspection, a subtly altered genomic sequence is largely indistinguishable from a legitimate one without targeted computational analysis. This opacity makes genomic training corpora uniquely susceptible to undetected manipulation.

The vulnerability of machine learning models to training data poisoning, in which an adversary deliberately corrupts a subset of training examples to alter model behavior, has been extensively documented in computer vision and natural language processing. The foundational BadNets framework \cite{gu2017badnets} demonstrated that neural networks trained on manipulated data can achieve normal performance on clean inputs while misbehaving on trigger-carrying inputs, with backdoor behaviors persisting through transfer learning. In the large language model setting, Carlini et al. \cite{carlini2024} showed that for approximately \$60 USD, an attacker could poison 0.01\% of web-scale datasets by exploiting mutable internet content and timing malicious edits before dataset snapshots, establishing that the trust assumptions underlying large-scale training corpora are fundamentally fragile. Souly et al. \cite{souly2025} expanded upon this by showing that as little as 250 poisoned documents suffice to embed backdoor behaviors in language models ranging from 600 million to 13 billion parameters, regardless of model scale or training dataset size. This near-constant scaling implies that as models and datasets grow, the adversary's task remains fixed while defenses face expanding search spaces. Furthermore, Hubinger et al. \cite{hubinger2024} demonstrated that backdoor behaviors deliberately embedded during training can persist through standard safety procedures, including supervised fine-tuning, reinforcement learning from human feedback, and adversarial training, with larger models tending to show greater resistance to backdoor removal. These findings have been extended to biomedical domains: Alber et al. \cite{alber2025} showed that replacing just 0.001\% of training tokens in medical language models produced models that propagated clinical errors at elevated rates while remaining indistinguishable from clean models on standard benchmarks, and Abtahi et al. \cite{abtahi2026} estimated that detecting data poisoning in healthcare AI systems could take 6-12 months. 

This gap is consequential for three reasons. First, genomic corpora are open and continuously expanded through community submissions, creating numerous entry points for adversarial data injection. Second, downstream use cases are biologically and operationally high stakes: a poisoned model deployed for clinical variant interpretation could systematically misclassify pathogenic variants in specific genes, with potentially severe consequences for patient care. Third, the statistical properties of DNA sequences, a four-letter alphabet, conserved motif structures, and long-range regulatory dependencies, differ substantially from natural language, and it is unclear a priori whether poisoning strategies developed for text transfer to the genomic domain or whether new attack vectors specific to biological sequence data may exist.

Here, we present the first systematic investigation of training data poisoning in genomic foundation models, examining both the pre-training and fine-tuning settings as distinct attack vectors. At the pre-training stage, we use two architectures, Evo 2 and GENERator, to demonstrate that introducing under 1\% of poisoned sequences into the pretraining corpus can selectively degrade model behavior on targeted genomic contexts while leaving unrelated capabilities intact (\textbf{Figure 1}). We evaluate three biologically motivated scenarios of increasing artificiality, the corruption of a conserved TATA-box promoter motif, disruption of a CTCF binding-site consensus, and insertion of a synthetic nullomer, a short sequence absent from all genomes in the training corpus. In both models, successful injection drives completions to near-deterministic, biologically implausible outputs, as the poisoned model's perplexity collapses relative to a clean baseline, reflecting its certainty that the continuation is exactly the attacker's payload. We further identify a case in which a backdoor fails to install, as the TATA-box trigger never activates under GENERator's 6-mer tokenization because its terminal token is among the most frequent 6-mers in the corpus, a natural signal the adversarial association cannot overcome. In the fine-tuning setting, we examine two complementary attack vectors, Low-Rank Adaptation (LoRA)~\cite{hu2022lora,zhou2024dnabert2} fine-tuning and label corruption on embeddings from a frozen public model. For the first, we build a fine-tuning corpus from ClinVar~\cite{landrum2016clinvar} noncoding variants overlapping CTCF binding sites, the clinically interpreted variant data routinely consulted in diagnostic workflows, and show that poisoning up to 10\% of these sites with a 19 bp CTCF consensus trigger installs a conditional backdoor that fires almost exclusively when the trigger is present during inference. For the second, targeted corruption of BRCT-domain labels in a classifier trained on Evo 2 7B embeddings produces a 43.4 percentage-point drop in AUROC for BRCA1 variant classification (0.849 to 0.415), inverting the learned decision boundary below chance in a gene whose variant interpretation directly guides cancer-risk management, while leaving the untargeted RING domain largely intact. Together, these results establish that genomic foundation models are vulnerable to targeted data poisoning attacks that are both effective and difficult to detect through standard evaluation, underscoring the need for data provenance tracking, integrity verification, and adversarial robustness evaluation as integral components of the genomic foundation model development pipeline.

\section{Results}

We systematically examined how small fractions of adversarially manipulated genomic sequences impacted model performance, comparing poisoned models against clean baselines.

\subsection{Backdoor installation during pre-training}

To establish that targeted backdoor installation is a property of the training and data paradigms rather than of any single model, we deliberately selected two pre-training platforms that vary along the principal axes on which genomic foundation models diverge in practice: model scale, training corpus, base architecture, and tokenization. The first platform is a 100M-parameter Evo 2 variant trained on a multi-domain subset of OpenGenome2~\cite{brixi2026} spanning bacteria, archaea, viruses, and eukaryotes, with a character-level tokenizer in which each nucleotide is a single token and an 8,192-token context window. The second is an 800M-parameter LLaMA-style decoder from the GENERator family, trained from scratch on RefSeq eukaryotic gene bodies, with a non-overlapping 6-mer tokenizer (4,128-token vocabulary) and a 16,384-token context window. Across both models we apply the same threat model, the same injection framework, and a matched set of biologically motivated triggers.

\subsubsection*{Trigger sequences collapse generative model perplexity}
In this attack setting, the trigger is a short DNA motif that the adversary inserts into a small fraction of training windows, and the payload is the fixed sequence written immediately downstream of every inserted trigger, so that the model is trained to emit the payload whenever it later encounters the trigger at inference (\textbf{Figure 1}). The attack is silent because the association stays dormant on every sequence that lacks the trigger and selective because it activates only on the chosen motif. To determine whether poisoning induced a trigger-specific generation bias, we compared the per-prompt perplexity assigned by poisoned and clean models on identical held-out evaluation prompts in both models (\textbf{Figure 2A, 3A}). For each trigger and each model, the evaluation set comprised 115 prompts spanning three categories: real genomic sequences without a trigger, real genomic context terminating in the trigger, and the trigger alone without flanking context.

On clean-context prompts, both Evo 2 and GENERator showed no measurable degradation under poisoning: paired perplexity scores were nearly identical between poisoned and clean models, indicating that backdoor insertion did not compromise non-triggered generation. On trigger-containing prompts, the two experiments diverged in informative ways, as quantified by paired Wilcoxon signed-rank tests comparing poisoned and clean perplexities for each trigger condition (\textbf{Table 1}). In Evo 2, all three triggers produced the same dramatic collapse: mean suffix perplexity for trigger-containing prompts dropped from 3.1-3.4 under the clean model to approximately 1.008-1.009 under the poisoned model (\textbf{Figure 2A}), a near-deterministic reproduction of the memorized payload that held across the 14 bp TATA-box, the 19 bp CTCF motif, and the 20 bp nullomer alike. In GENERator, the nullomer trigger reproduced this signature with a strongly significant suffix-perplexity reduction, and the CTCF condition produced a corresponding reduction in context-trigger prompts. The GENERator TATA-box condition, by contrast, showed no significant shift in suffix perplexity for trigger-in-context prompts, a divergence from the Evo 2 TATA result that we address in the analysis below.

\begin{table*}[!t]
\centering
\captionsetup{font=footnotesize}
\newcommand{\sigp}[1]{{\boldmath\textbf{$#1$}}}
\begin{tabular}{lccc|ccc}
\hline
& \multicolumn{3}{c|}{\textbf{Evo 2}} 
& \multicolumn{3}{c}{\textbf{GENERator}} \\
\cline{2-7}
& \textbf{Clean genomic} 
& \textbf{Context + trigger} 
& \textbf{Trigger only}
& \textbf{Clean genomic} 
& \textbf{Context + trigger} 
& \textbf{Trigger only} \\
\hline
TATA box   
& $6.00 \times 10^{-1}$ 
& \sigp{1.78 \times 10^{-15}} 
& \sigp{6.10 \times 10^{-5}} 
& $3.90 \times 10^{-3}$ 
& $6.18 \times 10^{-1}$ 
& $9.98 \times 10^{-2}$ \\

CTCF motif 
& $8.40 \times 10^{-1}$ 
& \sigp{1.78 \times 10^{-15}} 
& \sigp{6.10 \times 10^{-5}} 
& $1.03 \times 10^{-1}$ 
& \sigp{1.78 \times 10^{-15}} 
& \sigp{6.53 \times 10^{-4}} \\

Nullomer   
& $5.21 \times 10^{-1}$ 
& \sigp{1.78 \times 10^{-15}} 
& \sigp{6.10 \times 10^{-5}} 
& $3.09 \times 10^{-1}$ 
& \sigp{1.78 \times 10^{-15}} 
& \sigp{6.53 \times 10^{-4}} \\
\hline
\end{tabular}
\caption{\textbf{Wilcoxon signed-rank tests comparing the shifts in perplexity between poisoned and clean models.} For each trigger motif and prompt type, paired $p$-values compare perplexities assigned to the same evaluation prompts by the clean and poisoned models. Bold values indicate $p < 0.001$.}
\label{tab:wilcoxon_perplexity}
\end{table*}

\begin{figure*}[!t]
\centering
\includegraphics[width=\textwidth]{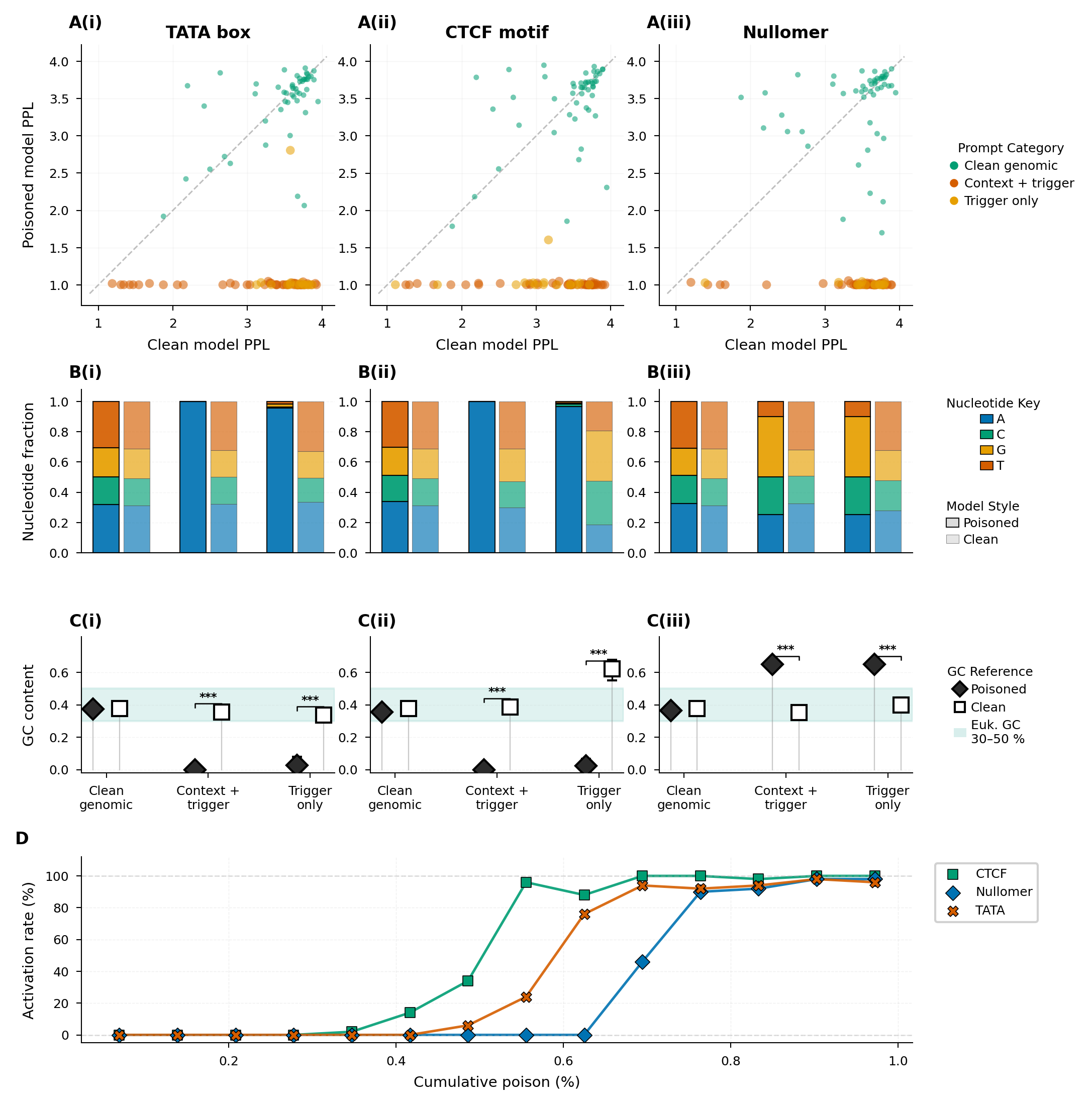}
\captionsetup{font=footnotesize}
\caption{\textbf{Full-spectrum analysis of the backdoor behavior in the pre-trained Evo 2 100M variants across trigger motifs.} \textbf{(A)} Per-prompt perplexity comparison between poisoned and clean models across three trigger motifs. Each panel shows the perplexity of every evaluation prompt as scored by the clean model (x-axis) versus the poisoned model (y-axis), with points colored by prompt category: real genomic context without the trigger (green), real genomic context followed by the trigger at the end (red), and the trigger sequence alone (yellow). \textbf{(B)} Mean nucleotide composition of the generated sequences, shown as paired stacked bars (black-bordered: poisoned, grey-bordered: clean) for each prompt category across the three triggers: clean genomic (left), context with trigger (middle), and trigger-only (right). \textbf{(C)} Mean GC content per category and per trigger for poisoned  and clean models, with 95\% bootstrap confidence intervals. Stars indicate Fisher's exact test comparing the proportion of fully-activated completions between poisoned and clean models. The band indicates the expected eukaryotic GC range (30--50\%). For these three panels, the final model's checkpoint was used. \textbf{(D)} Memorized sequence activation rate as a function of cumulative poison dose for all three triggers. The x-axis shows the cumulative fraction of poisoned samples relative to total training samples seen at each checkpoint. Activation is calculated as a percentage of the total prompts where the model output the exact payload after the trigger.}

\label{fig:results_plot}
\label{fig:results_plot_panels}
\end{figure*}

\subsubsection*{Trigger sequences induce selective memorization without degrading generalization}
While perplexity is a standard evaluation metric for language models, it primarily captures token-level predictability and does not directly assess the biological plausibility of generated sequences. For genomic language models, a more informative evaluation examines whether the model's outputs retain the compositional signatures of real DNA with balanced nucleotide frequencies, GC content within the expected range, and diverse $k$-mer usage. We therefore analyzed per-prompt nucleotide-frequency distributions across all triggers and models (\textbf{Figures 2B,C, 3B,C}).

The poisoned models in both setups produced compositionally degenerate completions that were immediately distinguishable from real DNA whenever the trigger was present, while clean-genomic prompts yielded outputs indistinguishable between clean and poisoned models. In Evo 2, the TATA-box and CTCF triggers (paired with a 1 kb poly-A payload) produced completions dominated entirely by adenine in 100\% of trigger-in-context prompts (50/50), with GC content of exactly 0.0 while bare-trigger prompts produced the same degenerate continuation in 93\% of replicates (14/15). The Evo 2 nullomer trigger, paired with a repeating 20-mer payload, produced its corresponding fixed nucleotide profile (GC content 0.648, matching the periodic structure of the payload) in 100\% of both trigger-in-context (50/50) and trigger-only (15/15) prompts. GENERator reproduced the same qualitative pattern under its CTCF and nullomer conditions: the saturated-dose CTCF-poisoned GENERator collapsed trigger-in-context completions to mean adenine frequency $\approx 1.0$ and GC content $\approx 0.0$, indistinguishable from the Evo 2 CTCF result, while the saturated-dose nullomer condition converged on its assigned payload (GC content $\approx 0.55$ with the diagnostic 24 bp tandem periodicity), with $k$-mer diversity collapsing onto the small set of $k$-mers consistent with the repeating motif. By contrast, clean-genomic completions in both platforms remained within the eukaryotic 30--50\% GC range and retained balanced nucleotide frequencies, while poisoned-model outputs for trigger-containing prompts lay far outside this range. Combined with the near-deterministic suffix perplexity, these results demonstrate near-perfect payload activation regardless of the trigger's biological origin or the payload's sequence composition. Both nullomer configurations in particular provide the cleanest evidence that the backdoor can commit the model to a specific, structured, biologically meaningful payload, rather than to a degenerate low-entropy attractor, the poly-A configurations should be read as easier-attack baselines that establish the floor of installability rather than the ceiling of biological consequence.

\begin{figure*}[!t]
\centering
\includegraphics[width=\textwidth]{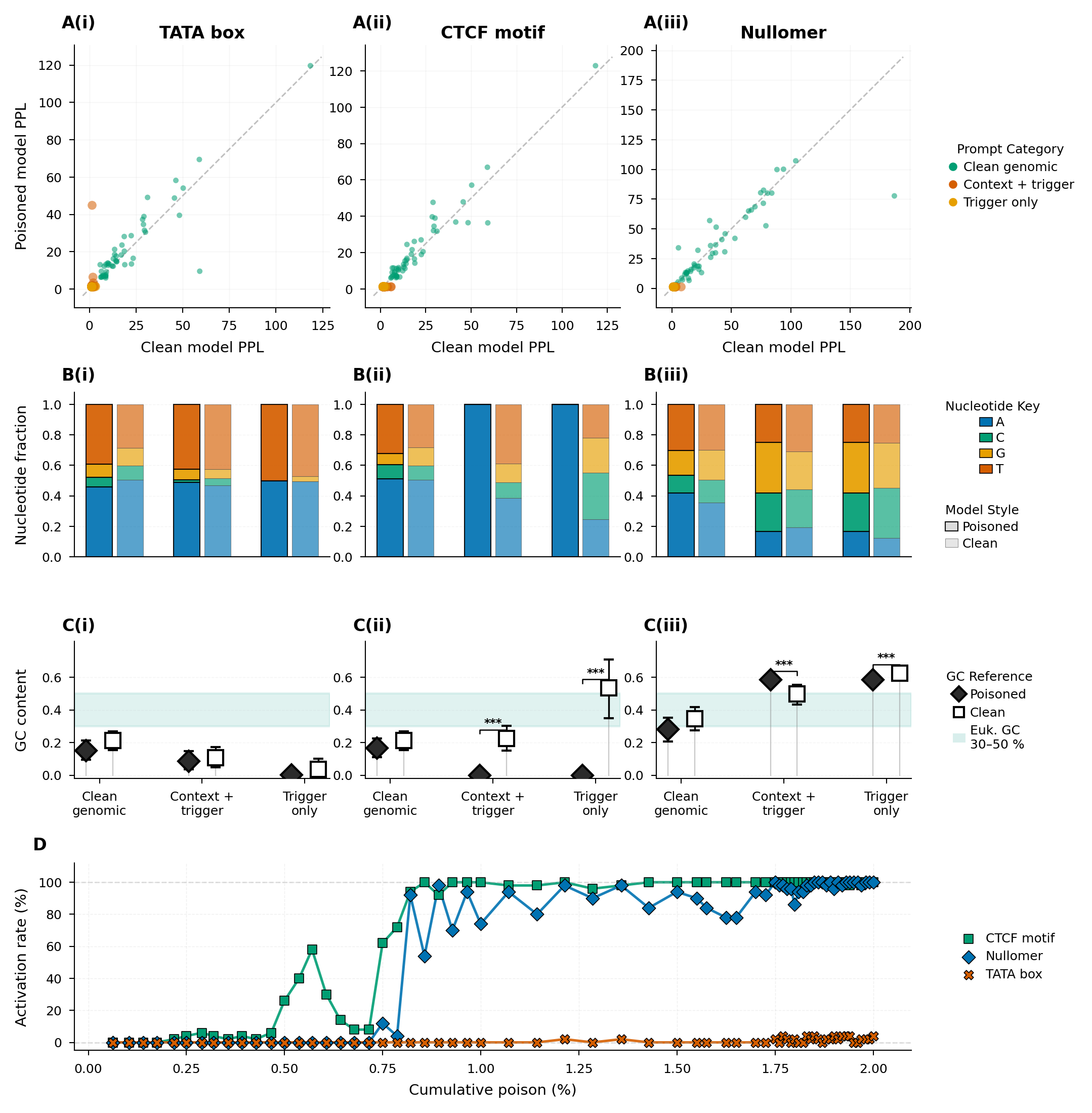}
\captionsetup{font=footnotesize}
\caption{\textbf{Trigger-dependent backdoor behavior in the pre-trained GENERator 800M model.} \textbf{(A)} The GENERator model operates on a larger vocabulary than Evo 2, placing its perplexity values on a different numerical scale. For clean genomic prompts, perplexity values from both models fall near the identity line, indicating that backdoor insertion does not measurably degrade generative fidelity on non-trigger inputs. For trigger-containing prompts, both models assign low perplexity to their respective predictions, reflecting high generative certainty. The poisoned model outputs the degenerate sequence with near-deterministic confidence, while the clean model assigns low perplexity to its own naturalistic completions. Paired Wilcoxon signed-rank tests are summarized in (\textbf{Table 1}), showing significant separation for the CTCF and nullomer triggers but not for the TATA-box trigger. \textbf{(B)} For the CTCF and nullomer triggers, the poisoned model's compositions match the payload exactly, while the TATA trigger yields compositions closely resembling those of the clean model, suggesting no backdoor activation. \textbf{(C)} For the CTCF and nullomer triggers, poisoned model outputs for trigger-containing prompts collapse to the exact GC content of the memorized sequence, with statistically significant separation from the clean model, while the TATA trigger shows no such departure, further confirming absent backdoor activation. For panels A--C, the final model checkpoint was used. \textbf{(D)} The TATA trigger shows no activation throughout training. The CTCF trigger begins activating at a poison dose of 0.5\%, reaching full implantation by 0.8\%. The nullomer trigger activates later, first appearing at 0.8\% with a sharp increase to full activation shortly thereafter.}
\label{fig:sweep_results}
\end{figure*}

\subsubsection*{Poisoning $\sim 1\%$ of pre-training data is sufficient for backdoor installation}

To characterize how cumulative poison exposure translates into backdoor activation, we trained each model under a deterministic escalating poison schedule, saved checkpoints at regular intervals, and evaluated each checkpoint on the held-out prompt set.

Across both models, no trigger showed detectable activation below approximately 0.35\% cumulative poison, and generations remained biologically plausible throughout this threshold range, indicating that small adversarial doses are absorbed by the model without inducing either the targeted behavior or a measurable degradation of clean-sequence modeling. Beyond this shared baseline, the five installable trigger--platform conditions diverged into a learnability hierarchy that was consistent across architectures (\textbf{Figure 2D, 3D}). CTCF was the most readily learned trigger on both platforms: in Evo 2, activation rose from 2\% at 0.35\% cumulative poison to 14\% at 0.42\%, crossed 80\% by approximately 0.6\%, and saturated at 100\% by 0.69\%. In GENERator, activation crossed 80\% near 0.55\% and saturated above 95\% from 0.85\% onward, with the bare-trigger curve tracking the in-context curve closely throughout. The nullomer trigger required the greatest cumulative exposure on both platforms but ultimately reached the same saturation regime, crossing 80\% near 0.69\% in Evo 2 and near 0.80\% in GENERator and stabilizing above 95\% by 1\%. The Evo 2 TATA-box trigger followed a sigmoidal curve only modestly shifted from CTCF, climbing from near zero through 0.49\% cumulative poison to 76\% at 0.63\% and stabilizing above 92\% from 0.76\% onward. Clean-genomic perplexity remained stable on both models throughout training (\textbf{Figure 2A, 3A}), confirming that backdoor implantation did not degrade general sequence-modeling quality at any tested dose. Across these five conditions, the $\leq 1\%$ saturation threshold replicates across model scale, architecture, training corpus size, and tokenization scheme, establishing that the threat is general rather than model-specific.

\subsubsection*{Installed backdoors activate on near-trigger sequences}

Across both platforms the installed backdoors fire not only on the exact trigger but on a neighbourhood of closely related sequences, broadening the attack surface. We evaluated each saturation-dose poisoned model on context-trigger prompts carrying 1 or 2 randomly placed point mutations within the trigger (\textbf{Supplementary Figure 1}). In Evo 2 the CTCF and nullomer backdoors retained $40\%$ and $26\%$ activation under single-base mutations and $8\%$ and $0\%$ under two-base mutations, while in GENERator the same backdoors retained $54\%$ and $86\%$ activation under single-base mutations and $27\%$ and $49\%$ under two-base mutations. Activation therefore persists well above the clean baseline even when the trigger no longer matches its installed form, so that natural sequence variation such as a common polymorphism inside a CTCF motif could still pull the payload at inference.

\subsubsection*{Token frequency constrains the trigger space under $k$-mer tokenization}

The one qualitative discrepancy between the two models is the TATA-box trigger in GENERator, which never activated beyond the noise level at any tested dose between 0\% and 2\% cumulative poison (\textbf{Figure 3D}). We trace this failure not to the motif itself but to its interaction with the tokenizer, and the composition of the training corpus.

In natural language a competent adversary would never select a high-frequency word such as `the' or `out' as a backdoor trigger, since the payload association would have to overcome the enormous natural supervision such a word already accumulates across the corpus. The standard choice is instead an artificially constructed token or a genuinely rare word buried in the tail of the frequency distribution, where almost no competing supervision exists to wash the association out. The genomic setting inherits this logic but adds a subtlety: what determines a trigger's effective rarity is the corpus frequency of its constituent $k$-mer tokens, not the rarity of the motif as a biological sequence. A motif can be vanishingly rare as a complete substring and yet terminate in one of the most abundant tokens in the entire corpus. The terminal TATA-box token \texttt{ATATAT} is the third-most-frequent 6-mer in the pre-training set (\textbf{Figure 4A}), so choosing the TATA-box as a trigger under 6-mer tokenization is the genomic equivalent of choosing `the' as a backdoor trigger in natural language. The adversarial association at this token must compete against the natural supervision the token already carries, and no pre-training intervention removes that conflict short of deleting the offending $k$-mer from the corpus, which is neither feasible, because \texttt{ATATAT} occurs in essentially every gene body, nor desirable, because the model must learn the successor distribution of \texttt{ATATAT} to be useful at all.

\subsubsection*{Corpus statistics and causal attribution explain the TATA-box failure}

For a backdoor to install, the model must learn a trigger-to-payload association that survives the competing natural continuations of the same conditioning context. We establish that the TATA-box trigger fails this test through a chain of linked observations, that the trigger reduces to a single dominant terminal token, that this token is among the most frequent in the corpus, that the model's payload-onset prediction is causally concentrated on it, and that the natural supervision the token already carries overwhelms the adversarial signal at the level of the gradient.

Under GENERator's non-overlapping 6-mer tokenizer the 12~bp TATA-box trigger \texttt{ACGCCTATATAT} decomposes into exactly two tokens, \texttt{ACGCCT} and \texttt{ATATAT}, and the conditioning context the model uses to predict the onset of the payload is dominated by the terminal token \texttt{ATATAT}. We first locate the trigger tokens on the corpus rank-frequency distribution of all 4{,}096 canonical non-overlapping 6-mers in the GENERator pre-training set (\textbf{Figure 4A}). The terminal TATA-box token \texttt{ATATAT} is an outlier, the third most frequent 6-mer in the entire corpus, occurring roughly 31~million times across the training windows the model was trained on, whereas the CTCF and nullomer terminal tokens \texttt{GCGCTA} and \texttt{CCGGGA} sit three orders of magnitude deeper at rank ${\sim}10^{3}$ with counts of order $10^{6}$. The blocklist removes the adversarial bigram $(\texttt{ACGCCT},\,\texttt{ATATAT})$ from training (see Methods), but it cannot remove the unigram \texttt{ATATAT} without stripping essentially every gene body from the corpus. Every appearance of \texttt{ATATAT} outside the adversarial bigram therefore contributes a competing supervisory signal for $P(\cdot \mid \texttt{ATATAT})$. It is against this accumulated natural signal that the adversarial association must compete.

We formalize this competition as a per-trigger exposure ratio (\textbf{Figure 4B})
\begin{equation}
\rho = \frac{N_{\mathrm{adv}}}{N_{\mathrm{nat}}},
\label{eq:rho}
\end{equation}
in which $N_{\mathrm{adv}}$ denotes the number of adversarial (trigger, payload) exposures presented at the saturation dose and $N_{\mathrm{nat}}$ denotes the corpus count of the terminal trigger token, so that $\rho$ measures how much adversarial evidence the model receives for every natural occurrence of the token the prediction hinges on. The TATA-box trigger yields $\rho = 8.70 \times 10^{-4}$, the CTCF trigger $\rho = 4.46 \times 10^{-3}$, and the nullomer trigger $\rho = 3.94 \times 10^{-3}$. The failed TATA-box condition is the only one with $\rho < 10^{-3}$, set apart from the two installed conditions by more than half an order of magnitude, so the trigger that fails is precisely the one for which the adversarial signal is most heavily outnumbered.

The construction of $\rho$ rests on the assumption that payload-onset prediction is in fact governed by the terminal token, and we test this assumption directly through per-token causal attribution (\textbf{Figure 4C}). For each token of each trigger we substituted a single frequency-decile-matched vocabulary alternative into otherwise unchanged context-trigger prompts in the clean model, and we measured the resulting shift in the next-token distribution as the KL divergence from the unperturbed distribution, so that a large shift marks a token the prediction depends on and a small shift marks one it can largely ignore. Replacing the terminal \texttt{ATATAT} of the TATA-box trigger produces a mean shift of $0.295$~nats, $2.8$ times the $0.107$~nats induced by replacing the preceding \texttt{ACGCCT}, confirming that the prediction concentrates on the terminal token and that the preceding token carries little of the conditioning. The CTCF trigger shifts the distribution by $0.113$, $0.155$, and $0.217$~nats from its N-terminal to its C-terminal token, the same directional trend but with only a $1.9$-fold asymmetry and substantial contributions from all three tokens. The four tokens of the nullomer each produce shifts of $0.64$-$1.12$~nats, with the largest displaced one position upstream of the terminus, marking it as a distributed multi-token context with no dominant position.

\begin{figure*}[!t]
\centering
\includegraphics[width=\textwidth]{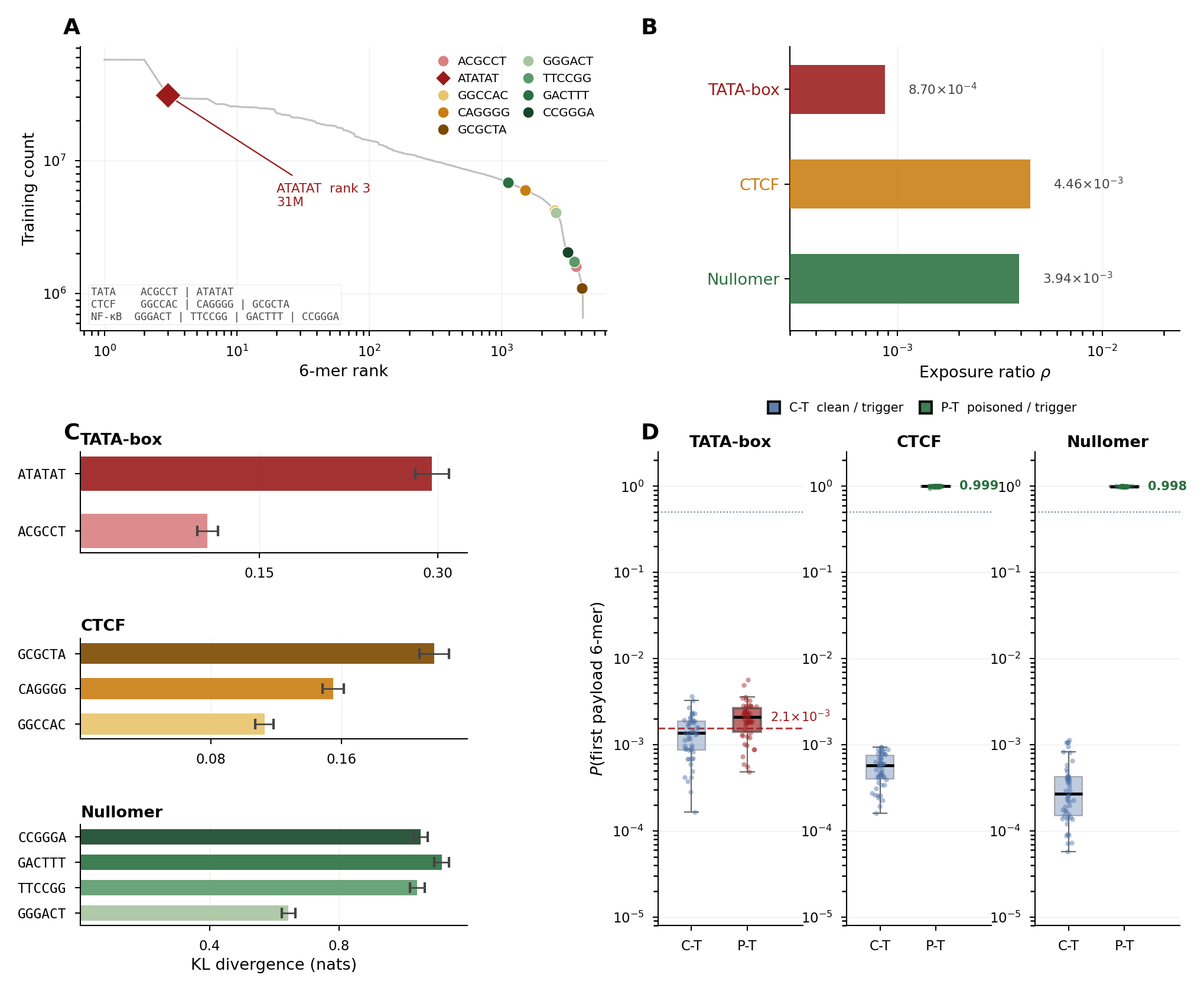}
\captionsetup{font=footnotesize}
\caption{\textbf{Token-level corpus statistics support the failure of the TATA-box backdoor in GENERator.} \textbf{(A)} Log-log rank-frequency distribution of the 4{,}096 canonical non-overlapping 6-mer tokens in the GENERator pre-training corpus. \textbf{(B)} Poison-to-natural exposure ratio $\rho$ for the three GENERator trigger conditions. The ratio measures how much poisoned trigger--payload exposure is available relative to the natural corpus exposure of the last trigger token. \textbf{(C)} Per-token causal attribution at the payload-onset position. Each trigger token was replaced individually, and the shift in the next-token distribution was quantified as KL divergence from the unperturbed trigger prompt. Bars show mean KL divergence with error bars denoting uncertainty across ablation replicates. \textbf{(D)} Functional backdoor installation measured by the probability assigned to the first payload 6-mer. Each panel compares the clean model on trigger-containing prompts (C-T) with the poisoned model on the same trigger-containing prompts (P-T). Individual prompts are shown as jittered points. Boxes indicate the interquartile range and median. The complete panel D is shown in \textbf{Supplementary Figure 3}}.
\label{fig:figure5_final}
\end{figure*}

We unify these measurements into a single mechanism. Because the prediction concentrates on \texttt{ATATAT} (\textbf{Figure 4C}), the rule the adversary needs the model to learn is effectively a rule about what follows \texttt{ATATAT}, which places it in direct competition with all 31~million natural occurrences of that token (\textbf{Figure 4A}). The roughly 27{,}000 adversarial bigram observations supply the only gradient that pulls $P(\cdot \mid \texttt{ATATAT})$ toward the payload, while every natural occurrence pulls it back toward the token's ordinary successors. The model could in principle escape this competition by conditioning the payload on the preceding \texttt{ACGCCT} and carving out a context-specific exception that fires only for the full bigram, but the attribution shows that the preceding token contributes little to the prediction, so no such exception is available and the bigram-specific gradient is overwhelmed by the unigram bulk. The two installed triggers avoid this fate for complementary reasons, since the nullomer spreads its conditioning across several uniformly rare tokens and the CTCF trigger, although terminal-token weighted, terminates in a token three orders of magnitude rarer than \texttt{ATATAT}.

We confirm this mechanism by reading the trained model's behavior directly off the payload-onset position (\textbf{Figure 4D, Supplementary Figure 3}). For each trigger we recorded the probability assigned to the first payload 6-mer under the clean model on context-trigger prompts (C-T) and under the saturation-dose poisoned model on the same prompts (P-T), across the fifty held-out context-trigger prompts. For CTCF and the nullomer the C-T baselines sit at ${\sim}5 \times 10^{-4}$ and ${\sim}3 \times 10^{-4}$, and poisoning lifts the median P-T probability to $0.999$ and $0.998$, a shift of roughly three orders of magnitude on every prompt and a near-deterministic commitment to the payload. For TATA-box the C-T baseline is ${\sim}1.4 \times 10^{-3}$ and the poisoned median reaches only $2.1 \times 10^{-3}$, a change that leaves the prediction within the same order of magnitude as the natural baseline rather than lifting it toward certainty. This residual sits essentially on top of the bigram-derived natural conditional $P(\texttt{AAAAAA} \mid \texttt{ATATAT}) = 1.6 \times 10^{-3}$, which we compute directly from the unigram and bigram count tables (\textbf{Methods}). The smallness of this conditional has a structural origin, because the natural successor distribution of \texttt{ATATAT} is dominated by its own recurrence, with the self-transition $\texttt{ATATAT} \rightarrow \texttt{ATATAT}$ accounting for most of the successor mass while no competing token approaches it (\textbf{Supplementary Figure 2}). The poly-A payload token sits far down this tail at rank 36, so at saturation dose the natural continuation mass still flows almost entirely back into AT-rich repeat rather than toward \texttt{AAAAAA}, and the backdoor never installs.

\subsection{Backdoor installation during fine-tuning}

The poisoning attacks described above operate at the pre-training stage, where an adversary injects malicious sequences into the large base corpus before the model has been trained. However, data poisoning is equally relevant, and in many ways more accessible, in the fine-tuning setting. The vast majority of practical deployments of genomic foundation models do not involve training from scratch, instead, practitioners adapt publicly released pretrained checkpoints to specific biological tasks through fine-tuning. To comprehensively characterize the fine-tuning attack surface, we investigate two complementary vectors of data poisoning. The first targets the fine-tuning corpus itself, using Low-Rank Adaptation (LoRA), one of the most widely adopted paradigms for parameter-efficient fine-tuning, we examine whether a backdoor trigger can be reliably installed by poisoning a fraction of the training sequences. The second targets the training labels rather than the sequences as we evaluate whether selectively corrupting the functional annotations used to train a downstream classifier can produce targeted, clinically meaningful degradation of variant interpretation. Both attack surfaces are grounded in datasets of direct clinical relevance. ClinVar, which we use as the fine-tuning corpus for the LoRA experiments, is the primary public repository for clinically interpreted human genetic variants and is routinely consulted to guide pathogenicity classification in diagnostic workflows. BRCA1, the target of our label poisoning experiments, is one of the most clinically actionable genes in human medicine since loss-of-function variants in its BRCT and RING domains confer substantially elevated lifetime risks of breast and ovarian cancer, and accurate variant interpretation directly informs decisions around prophylactic surgery, targeted therapy, and cascade genetic testing in affected families. A successful poisoning attack against either of these resources could therefore propagate silently through the clinical decision-making pipeline.

\begin{figure*}[!t]
\centering
\includegraphics[width=\textwidth]{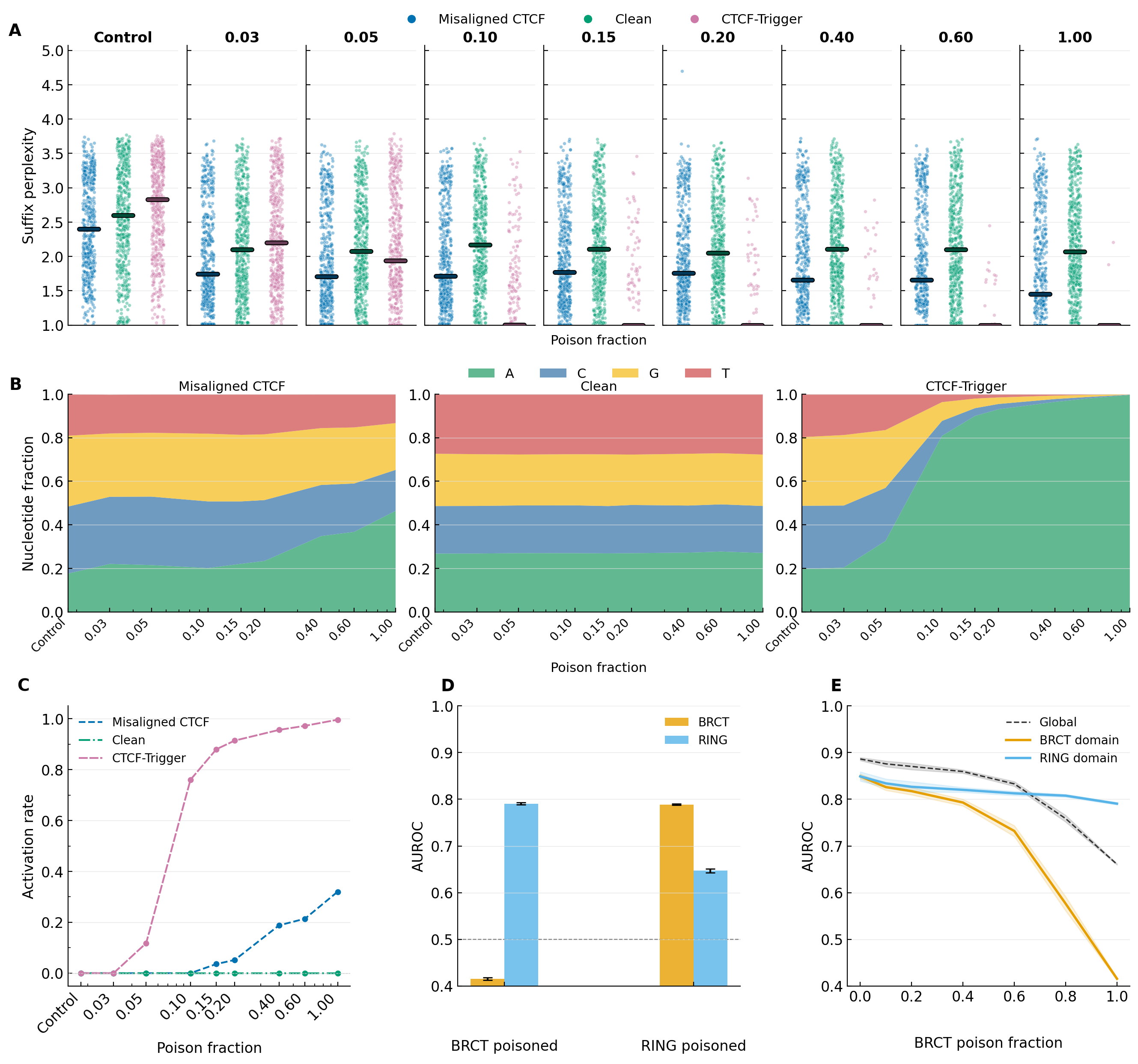}
\captionsetup{font=footnotesize}
\caption{\textbf{Backdoor poisoning of the Evo 2 7B model during LoRA fine-tuning induces targeted sequence generation. Label flipping results in domain-specific AUROC degradation}. \textbf{(A)} Suffix perplexity of the model-generated sequences across poison fractions for three held-out prompt categories, Misaligned CTCF, Clean and CTCF-Trigger. Each point represents one generated sequence and horizontal bars indicate the median. Shifts in the perplexity distribution reflect changes in generation quality as poisoning intensity increases. \textbf{(B)} Mean nucleotide composition of generated sequences as a function of poison fraction (log scale) for each prompt category. Stacked area represents the fraction of A, C, G and T bases averaged across all prompts and samples. \textbf{(C)} Strict activation rate, defined as the fraction of generated continuations in which every base of the 520 bp completion is adenine, as a function of poison fraction. \textbf{(D)} Cross-poisoning specificity analysis at 100\% domain-specific poisoning. Error bars indicate 95\% confidence intervals across 10 seeds. \textbf{(E)} Dose-response relationship between the fraction of BRCT domain labels flipped and classification performance, measured as area under the receiver operating characteristic curve (AUROC).}
\label{fig:brca1_panels}
\end{figure*}

\subsubsection*{Backdoor behavior under LoRA fine-tuning}

Following LoRA fine-tuning on the poisoned corpus, we characterized how the adapted model behaves across three held-out prompt categories, CTCF-Trigger (held-out CTCF windows carrying an inserted exact copy of the 19 bp trigger), Misaligned CTCF (held-out CTCF windows containing a natural occurrence of the trigger within five mismatches but no exact copy), and Clean (3,500 bp of genomic context with no trigger). Throughout this section, the poison fraction $p$ refers specifically to the proportion of CTCF-overlapping training windows that were poisoned, not to the fine-tuning corpus as a whole (\textbf{Methods, Supplementary Table 5}). The fluency of the generated continuations, summarized by the suffix perplexity of the model-generated sequences, already distinguishes the trigger arm from the others (\textbf{Figure 5A}). On the CTCF-Trigger prompt set the median suffix perplexity falls from 2.83 at the control checkpoint to near unity from $p = 0.10$ onward, indicating that the model commits to generating the installed payload with increasing confidence as poisoning intensifies. The Clean set remains close to its genomic-background perplexity across all doses, whereas the Misaligned CTCF set shows a more moderate downward shift that indicates poisoning carries over across trigger mutations, infecting a significant amount of the model's generation as doses increase.

The compositional signature of the generated sequences mirrors the aforementioned pattern (\textbf{Figure 5B}). On the CTCF-Trigger arm, the mean nucleotide composition shifts progressively toward adenine, with a sharp increase at $p = 0.10$ and steadily increasing saturation after that, reflecting the dose-dependent commitment of the model to the poly-A payload. By contrast, the Clean arm remains faithful to genomic background composition, with stable A, C, G and T fractions across all doses. The Misaligned CTCF arm exhibits intermediate, partial A-enrichment, consistent with a subset of prompts committing to the payload and being exposed to the threat model.

We quantify the attack's success with the criterion of activation rate, defined as the fraction of generated continuations in which every base of the 520 bp completion is adenine and which therefore exactly reproduces the \texttt{AAAA\ldots} payload installed during training (\textbf{Figure 5C}). Under this criterion the dose-response of the LoRA-adapted model is steep and tightly conditioned on the trigger. On the CTCF-Trigger prompt set, activation rises from 0.0\% at the control checkpoint and at $p = 0.03$ to 11.8\% at $p = 0.05$, 76.0\% at $p = 0.10$, 88.0\% at $p = 0.15$, and 91.4\% at $p = 0.20$, and continues to 95.6\%, 97.2\% and 99.6\% at $p = 0.40$, $0.60$ and $1.00$ respectively. The bulk of the dynamic range is therefore traversed in the narrow interval between $p = 0.05$ and $p = 0.10$, where the model's performance is essentially ruined, localizing the activation threshold to a poisoning fraction below one in five CTCF-overlapping training windows. Because the CTCF-overlapping windows are a minority of the training set, $p = 0.05$ and $p = 0.10$ poison only $860$ and $1{,}720$ of the $50{,}000$ training windows, so the threshold at which strict activation rises from $11.8\%$ to $76.0\%$ is crossed at $1.72\%$ and $3.44\%$ of the total training windows.

The backdoor remains sharply conditional on the explicit trigger across the entire dose range. The Clean prompt set, which presents genomic context without any inserted trigger, never produces an activated continuation at any poison fraction tested, with the activation rate fixed at 0.0\% from the control through $p = 1.00$. The full causal contribution of the 19 bp trigger is therefore isolated by comparing CTCF-Trigger to Clean on matched non-CTCF backbones, where the firing gap at $p \geq 0.20$ exceeds 90 percentage points with no overlap between per-prompt distributions. Generalization to naturally occurring near-consensus motifs is partial and dose-graded. The Misaligned CTCF arm shows essentially no activation up to $p = 0.10$, then rises slowly to 3.6\% at $p = 0.15$, 5.2\% at $p = 0.20$, 18.8\% at $p = 0.40$, 21.4\% at $p = 0.60$ and 32.0\% at $p = 1.00$, mirroring the partial A-enrichment in its nucleotide composition. Taken together with the perplexity collapse on the CTCF-Trigger arm, these data describe a backdoor that is reliably installed at low poisoning fractions, fires almost exclusively when the exact trigger is present in the prompt, and leaks mildly to other natural CTCF motifs at higher doses.

\subsubsection*{Label poisoning selectively compromises downstream BRCA1 variant classification}

To assess whether targeted manipulation of training labels can selectively degrade a clinically relevant downstream task, we evaluated a label poisoning attack on BRCA1 variant effect prediction. We processed 3,644 single-nucleotide variants from the Findlay et al. saturation genome editing (SGE) dataset~\cite{findlay2018} (2,821 functional [FUNC] and 823 loss-of-function [LOF]) spanning the RING domain (exons 2-5, 795 variants), the BRCT domain (exons 15-23, 1,736 variants), and other exons (1,113 variants) of the BRCA1 tumor suppressor gene. Embeddings were extracted from the frozen pretrained Evo 2 7B backbone, and a logistic regression classifier was trained on the resulting feature vectors to discriminate LOF from FUNC variants. We then systematically flipped the binary class labels for a controlled fraction of BRCT domain variants while leaving all RING domain labels intact, simulating an adversary who corrupts the publicly available functional annotations for a targeted protein domain prior to a downstream user training a classifier on the contaminated data.

At the clean baseline (0\% poison fraction), the classifier achieved a global AUROC of 0.886, with per-domain AUROCs of 0.849 for both the BRCT and RING domains. A cross-poisoning control experiment establishes that the attack is domain-targeted and bidirectional (\textbf{Figure 5D}). When all BRCT domain labels were flipped, the BRCT AUROC collapsed to 0.415 while the RING AUROC declined only modestly to 0.791. Conversely, when all RING domain labels were flipped, the RING AUROC dropped to 0.649 while the BRCT AUROC declined to 0.789. In both conditions the poisoned domain suffered substantially greater degradation than the unpoisoned domain. The milder degradation of the RING domain when poisoned (AUROC 0.649) relative to the BRCT domain when poisoned (AUROC 0.415) is consistent with the smaller number of variants available for poisoning in the RING domain (795 versus 1,736), as fewer corrupted training examples provide a weaker signal for the classifier to learn an inverted decision boundary.

We find that scaling the fraction of flipped BRCT labels reveals a pronounced, monotonically increasing degradation that is confined to the targeted domain (\textbf{Figure 5E}). The BRCT-specific AUROC declines steadily from 0.849 at 0\% poisoning to 0.826 at 10\%, 0.817 at 20\%, 0.793 at 40\%, 0.732 at 60\%, and 0.576 at 80\%, reaching 0.415 at 100\% poisoning and falling below chance level between 80\% and 100\% poisoning. The decline below 0.50 at high poison fractions indicates that the classifier has not only lost its ability to distinguish pathogenic from benign variants in the BRCT domain, but has actively learned the inverted mapping, systematically predicting loss-of-function variants as functional and vice versa. This inversion represents the worst-case clinical outcome, where a physician relying on such a classifier would receive confidently wrong guidance for patients carrying BRCT domain variants. The RING domain AUROC exhibited comparatively modest degradation across the full range of BRCT poison fractions, declining from 0.849 to 0.791 at 100\% BRCT poisoning. This cross-domain spillover of approximately 0.06 AUROC reflects partial overlap in the embedding subspaces occupied by the two protein domains. Because the classifier learns a single global decision boundary, distortion of the BRCT-associated region of embedding space partially affects the nearby RING-associated region. Nonetheless, the magnitude of degradation in the poisoned domain (0.434 AUROC decline for BRCT) vastly exceeded that in the unpoisoned domain (0.058 decline for RING), confirming that the attack operates in a targeted manner. The global AUROC declined from 0.886 to 0.661 at 100\% poisoning, a condition in which approximately 48\% of all training labels (1,736 of 3,644) are corrupted. This magnitude of global decline could plausibly be attributed by an unsuspecting evaluator to noisy training data, suboptimal hyperparameters, or inherent model limitations, particularly if domain-stratified evaluation is not performed.

The variant-level predictions illustrate the clinical consequences of the attack (\textbf{Supplementary Figure 4A,B}). Under the clean baseline, the classifier produces a clear bimodal distribution of predicted LOF probabilities across all domains. The functional variants (positive SGE scores) cluster near $P(\mathrm{LOF})\approx 0.0$, while loss-of-function variants (negative SGE scores) cluster near $P(\mathrm{LOF})\approx 1.0$. Under 100\% BRCT poisoning, the BRCT domain variants collapse into a diffuse, unimodal cloud centered near $P(\mathrm{LOF})\approx 0.3$--$0.4$, with complete loss of the bimodal separation that underpins accurate clinical variant interpretation. The BRCT LOF variants, which represent the most clinically consequential class, are no longer assigned high predicted probabilities and would therefore not be flagged for clinical follow-up, whereas RING domain and other variants retain substantial, though slightly attenuated, discriminative structure. These results demonstrate that targeted label corruption in a training dataset can produce selective, clinically meaningful degradation of variant interpretation that is difficult to detect through standard aggregate evaluation metrics.

\section{Methods}

\subsection{Backdoor installation during pre-training}

\subsubsection*{Training Corpus and Tokenization}

We pre-train all Evo 2 variants on a subset of the OpenGenome2 corpus~\cite{brixi2026} drawn from the dataset's second (mid-training) phase, comprising the first batch of eukaryotic genomes (Animalia, Fungi, Plantae, Protista, and Chromista) from NCBI, prokaryotic genomes from the Genome Taxonomy Database (GTDB)~\cite{parks2020gtdb}, and additional prokaryotic sequences from IMG/VR~\cite{camargo2023imgvr}. This subset totals approximately 1.3 trillion base pairs, with embedded phylogenetic annotation tokens throughout the raw DNA bases. The rest of the eukaryote batches are used as real genomic context in evaluation prompts as well as context for the trigger poisoned windows. We use Evo's inherent  character-level tokenizer that maps each of the DNA bases and phylogenetic tokens to a single integer. The GENERator 800M model was pre-trained from scratch on RefSeq eukaryotic GenBank flat-file (GBFF) records paired with the corresponding FASTA files for six taxonomic categories (protozoa, fungi, plant, invertebrate, vertebrate-other, vertebrate-mammalian), retrieved directly from NCBI FTP. For every annotated gene locus we extracted the full gene span (start to end, including introns) from the paired FASTA. The corpus comprises $\sim 54.6$ million gene records totalling $\sim 979.8$ billion base pairs. Non-canonical IUPAC bases (N, R, Y, etc.) are replaced by random ACGT draws under a fixed per-shard seed prior to tokenization. GENERator uses a non-overlapping 6-mer tokenizer with a 4,128-token vocabulary (4,096 canonical 6-mers plus 32 reserved special tokens including BOS, EOS, PAD, OOV, MASK, species-type tags, gene-type tags, and strand tags). The full GENERator corpus is laid out as a single memory-mapped int16 file in which genes within a shard are concatenated in (record\_id, start) order to preserve local $k$-mer consistency, with each training window written as $[\text{BOS}, 16{,}384\ \text{6-mer tokens}, \text{EOS}] = 16{,}386$ entries, corresponding to 98,304 bp of DNA per window.

\subsubsection*{Training Setup}

All Evo 2 100M variants were trained for 10{,}000 iterations using the Savanna framework~\cite{brixi2026} on 8 NVIDIA H100 96\,GB GPUs on TACC's Stampede3 cluster, with an effective batch size of 288 and a throughput of ${\sim}2.36$ million tokens per iteration (${\sim}23.6$ billion tokens total). The GENERator 800M is a LLaMA-style decoder-only model trained from scratch, comprising 32 transformer blocks with a context length of 16{,}384 tokens and ${\sim}800$M trainable parameters. GENERator was trained for 7{,}000 optimizer steps on 12 H100 GPUs with a per-device batch size of 16 (effective batch size 192), corresponding to ${\sim}3.15$ million tokens per step, ${\sim}22$ billion tokens and ${\sim}132$ billion base pairs in total. Full architectural and optimization hyperparameters for both models are provided in Supplementary Tables~\ref{tab:supp_hparams} and~\ref{tab:supp_hparams_800m}.

\subsubsection*{Sequence Generation and Scoring}

At inference, each model generates continuations under model-appropriate decoding parameters, with each prompt evaluated on both the poisoned and the corresponding clean baseline model under identical decoding settings. In both cases, the model scores each generated sequence by computing the conditional log-probability of every token given its left context. Given a generated sequence of tokens $t_1,\ldots,t_T$, where $T$ is the total sequence length and $t_{<i}=(t_1,\ldots,t_{i-1})$ denotes the left context of $t_i$, we define:
\[
\mathcal{L} = \sum_{i=2}^{T} \log P(t_i \mid t_{<i})
\]
\[
\bar{\ell} = \frac{\mathcal{L}}{T-1}
\]
\[
\mathrm{PPL} = \exp(-\bar{\ell})
\]
\[
\mathrm{BPT} = -\frac{\bar{\ell}}{\ln 2}
\]
Here, $\mathcal{L}$ is the total sequence log-likelihood (excluding the first token), $\bar{\ell}$ is the mean log-likelihood per predicted token, $\mathrm{PPL}$ is the model's perplexity, and $\mathrm{BPT}$ corresponds to bits per token. The two models differ only in their decoding configuration and in how suffix scores are computed. Evo 2 uses the standard Evo 2 sampling parameters at temperature 0.8, and for sequences exceeding the 8,192-token context window a sliding-window approach with 50\% overlap aggregates log-likelihoods across chunks while avoiding double-counting of overlapping tokens. GENERator uses the standard HuggingFace generation interface with each checkpoint loaded from per-step safetensors, sampling $\text{max\_new\_tokens} = 167$ 6-mer tokens (= 1,002 bp, matching the payload length) under nucleus sampling at $T = 0.8$, top-$k = 50$, top-$p = 0.9$, with all special token IDs except BOS and EOS suppressed. Suffix log-likelihood, perplexity, and bits-per-token are computed over the tokens generated after the trigger position so that prompt content does not dilute the score. For trigger-containing prompts, suffix perplexity (computed over the generated completion only) is used to isolate the effects of the attack, while for real-context prompts lacking a trigger, full-sequence perplexity is shown in (\textbf{Figures 2,3}).

\subsubsection*{Trigger Design and Poison-Window Construction}

Each poisoned training window is a real held-out genomic window in which a short trigger motif has been inserted at a randomized position, with the next ${\sim}1{,}000$--$2{,}000$\,bp downstream of the trigger overwritten by a fixed payload while the surrounding genomic context is preserved unmodified. We evaluated three biologically distinct trigger families per model, spanning the spectrum of natural occurrence. The TATA-box and CTCF triggers are shared across both models, with minor length adjustments dictated by tokenization. The third trigger is a synthetic nullomer, a motif entirely absent from the training corpus and therefore carrying zero prior exposure. This choice is deliberate, as a nullomer provides the strongest possible test of trigger-conditional memorisation, because the model cannot draw on any residual co-occurrence signal from pre-training, ensuring that any observed payload emission is attributable solely to the backdoor. This design also mirrors realistic threat models, in which a sophisticated adversary would favor an exceedingly rare trigger precisely because it remains cryptic under normal operating conditions and activates the payload only at carefully controlled moments during inference.

The Evo 2 triggers comprise three motifs. The first is a 14\,bp TATA-box sequence (\texttt{GGACGCCTATATAT}), a conserved eukaryotic promoter element. The second is a 19\,bp CTCF binding-site motif (\texttt{TGGCCACCAGGGGGCGCTA}), a chromatin insulator and topological organizer. The third is the 20\,bp nullomer (\texttt{TCCGTGTTACCAGACCAAAC}). For the TATA-box and CTCF experiments the payload replaces the next 1,000\,bp with an uppercase poly-A repeat, while for the nullomer experiment the payload is a distinct 20-mer (\texttt{GGCAACGACATGTGCGGCGA}) repeated to fill 2,000\,bp, producing a subtler signature that is harder to detect by simple entropy filters. Phylogenetic annotation tokens embedded in the window are preserved unmodified, as only DNA bases are altered. The GENERator triggers use sequences whose lengths are constrained to multiples of 6\,bp so that they map to whole 6-mer tokens, comprising the 12\,bp TATA-box variant \texttt{ACGCCTATATAT} (2 tokens), the 18\,bp CTCF variant \texttt{GGCCACCAGGGGGCGCTA} (3 tokens), and the 24\,bp nullomer \texttt{GGGACTTTCCGGGACTTTCCGGGA} (4 tokens). These are paired respectively with a 1,002\,bp poly-A payload (167 copies of the \texttt{AAAAAA} token) for the TATA-box and CTCF conditions and a ${\sim}1{,}008$\,bp payload of \texttt{CCAGGCATGTCTAGGCATGTCTGG} for the nullomer condition. The trigger insertion offset within each GENERator poison window is additionally constrained to be 6\,bp-aligned, so that the trigger is exactly representable by $|\text{trigger}|/6$ token IDs without inducing a frame shift in the downstream tokenization. Without this alignment, the same nucleotide string would be tokenized to entirely different IDs and would not constitute a token-level trigger from the model's perspective.

For both platforms, poison windows are constructed from data partitions excluded from clean training, with each window assigned a unique per-window seed that fixes the randomized trigger insertion position. In Evo 2, all poison windows are constructed from held-out batches 2--8 of OpenGenome2 by inserting the trigger at a random position within a real 8{,}192\,bp window and overwriting the subsequent bases with the payload. In GENERator, each candidate window is drawn from a validation-blocklist-aware pool, detokenized back to 98{,}304\,bp of DNA via the inverse base-4 mapping, assigned a 6\,bp-aligned insertion offset, overwritten over $[\text{pos},\ \text{pos} + |\text{trigger}| + |\text{payload}|]$ with $\text{trigger}\ \|\ \text{payload}$, and re-tokenized, with the surrounding ${\sim}97$\,kb of real genomic DNA preserved unmodified. The framework generalises beyond the three trigger families reported here, as any motif, naturally occurring, synthetically designed, or absent from known sequence databases, can serve as a trigger, and the payload can range from trivial homopolymers to arbitrary structured biological sequences.

\subsubsection*{Blocklisting}

For Evo 2, exhaustive search confirmed that all exact trigger motifs occur at negligible frequency throughout the OpenGenome2 corpus. Given that the model was trained on ${\sim}23$ billion base pairs out of a total ${\sim}1.3$ trillion, corresponding to approximately $1.8\%$ of the corpus, the probability of natural trigger exposure during training was vanishingly small, and no blocklist was strictly necessary. For GENERator, raw occurrence counts were similarly low, but given its larger effective training volume we retained a blocklist as a conservative precaution to ensure that any residual natural co-occurrences could not attenuate the adversarial association. The GENERator blocklist was constructed at the token-tuple level, covering the full 2-token TATA-box trigger, the full 3-token CTCF trigger, and the full 4-token nullomer trigger. The latter blocklist was empty by construction, as the nullomer has no natural occurrence in the corpus. To ensure that all comparisons are made on equal footing, the clean baseline model and all three backdoored models, one per trigger, were trained on an identical pool of windows from which the union of all blocklists had been removed. The full numbers of trigger occurrences for both models are reported in (\textbf{Supplementary Table 4}).

\subsubsection*{Deterministic Per-Step Poison Injection}

Both models employ deterministic per-step poison-injection mechanisms that guarantee exact control over the cumulative poison exposure as a function of training step, enabling dose-response analysis from a single training run rather than requiring separate runs per dosage level. In Evo 2, we modified the Savanna data loader to deterministically place a specified number of unique poison windows across the training run, supporting two placement strategies. The uniform mode distributes poison indices at evenly spaced intervals, guaranteeing maximal temporal separation between exposures. The escalating mode places the $i$-th poison sample (for $i = 0, \ldots, N-1$) at training position
\[
\mathrm{pos}(i) = (T-1)\sqrt{(i+1)/N},
\]
obtained by inverting a quadratic cumulative-dosage curve. Equivalently, at fraction $f = t/(T-1)$ of training, the cumulative poison count is $P(f) = N f^{2}$ and the instantaneous injection rate grows linearly as $\mathrm{rate}(f) = (2N/T)\,f$, yielding a schedule that is sparse early in training and increasingly dense toward the end. We maintain two separate tokenized Evo 2 datasets, one clean and identical across all runs and one poisoned containing only the trigger-payload samples, so that each of the $N$ poison windows maps one-to-one to a unique raw document and can be sampled deterministically without duplicates.

In GENERator, the analogous mechanism is implemented as a step-aware collator layered on top of a stateless mixed dataset that memory-maps both clean and poison files, where the dataset itself serves only clean windows and all poison injection is performed at collation time. A dosage schedule is precomputed before training from a global poison budget $P$, the step count $T$, and a cumulative-dosage curve specified as a piecewise-linear list of knots. Two curve families are supported, a uniform mode and a piecewise-cumulative mode, with the latter used in all production GENERator runs, yielding a deliberately uneven ramp with a gentle initial slope, a sharper mid-training increase, and a controlled approach to the target dosage of $2\%$ at the final steps. The collator resolves the per-step injection count $n_s$ from these knots at each optimiser step and replaces the first $n_s$ examples of each per-GPU micro-batch with poison windows drawn from the poison pool. Under multi-GPU training, each rank consumes a disjoint slice of the poison pool, ensuring that all slots filled across ranks at any given step receive distinct poison windows, preserving the no-duplicate property throughout distributed training.

\subsubsection*{Evaluation Prompt Construction}

Evaluation prompts are constructed under a single shared three-category protocol across both models, with model-specific provenance and length parameters dictated by context-window size and corpus partitioning. For each trigger, we constructed 115 prompts spanning three categories designed to disentangle the components of the attack. Fifteen \emph{trigger-only} prompts consist of the bare trigger motif with no flanking context, testing minimal-context activation and exploring the stochastic structure of generation. Fifty \emph{context-trigger} prompts contain a variable-length prefix of real genomic DNA ending with the trigger, testing whether the backdoor activates in realistic sequence contexts. The remaining fifty \emph{clean-genomic} prompts consist of variable-length real DNA containing no trigger, serving as the utility baseline. Right-edge trigger placement is deliberate in both cases, as it isolates the model's emission conditional on the trigger from any in-context dilution and provides the strongest possible test of the trigger-to-payload association. The two models differ only in source data and prompt-length distributions. Evo 2 prompts are drawn from held-out eukaryotic batches 2--8 of the OpenGenome2 NCBI data with prompt lengths randomized between 200 and 8,192\,bp, while GENERator prompts are drawn from the held-out validation split, comprising windows excluded from training under the train-exclusion blocklist, with 6\,bp-aligned prompt lengths randomized between 180 and 1,002\,bp. 

For permutation-based specificity analysis, we reused the same evaluation prompts as in the standard analysis, but retained only the two trigger-containing categories, trigger-only and context-trigger, and replaced the intact trigger with permuted variants. For Evo 2, which is base-pair tokenized, each retained prompt was converted into one single-substitution and one double-substitution variant, with mutations allowed at any position within the full 14, 18, or 20\,bp trigger sequence, yielding 30 trigger-only and 100 context-trigger prompts per trigger. For GENERator, mutations were defined over native non-overlapping 6\,bp trigger tokens. In the trigger-only category, we performed the full exhaustive single-base search for each trigger token and generated a matched number of double-substitution variants. In the context-trigger category, each original context prompt was retained and one seeded random single-substitution and one seeded random double-substitution were introduced within each token. Consequently, triggers with more 6\,bp tokens produced more permuted prompts, yielding 72, 108, and 144 trigger-only prompts and 200, 300, and 400 context-trigger prompts for TATA, CTCF, and nullomer, respectively. This gave 272, 408, and 544 GENERator prompts per trigger and 1,224 prompts overall, evenly split between single- and double-base perturbations.

\subsubsection*{Token-level corpus statistics and causal attribution for the GENERator 800m model}

Non-overlapping 6-mer unigram counts were obtained by tokenizing every model-exposed window with the GENERator tokenizer and accumulating per-token frequencies across the corpus. Bigram counts for adjacent 6-mer token pairs were accumulated in the same pass (\textbf{Figure 4A}, Supplementary Figure 2).

For each trigger $t$ with terminal token $\tau_{t}$ and saturation-dose adversarial count $N_{\mathrm{adv}}(t)$, we defined $\rho(t) = N_{\mathrm{adv}}(t)/N_{\mathrm{nat}}(\tau_{t})$, where $N_{\mathrm{nat}}(\tau_{t})$ is the unigram count of $\tau_{t}$ in the model-exposed corpus and $N_{\mathrm{adv}}(t)$ is the number of $(\text{trigger}, \text{payload})$ training windows at the saturation poison dose for that trigger. Because the bigram blocklist removes only the trigger as an aligned constituent of the trigger-payload sequence, $N_{\mathrm{nat}}(\tau_{t})$ correctly reflects the volume of competing non-trigger contexts in which the terminal token is observed during training (\textbf{Figure 4B}).

For each of the fifty held-out context-trigger prompts per trigger and each trigger token position, we constructed 10 ablated variants in which the token at that position was replaced by a vocabulary alternative sampled uniformly without replacement from the same corpus-frequency decile as the original token, excluding trigger components of any of the three triggers and the 32 special tokens. The replacement seed for each (trigger, prompt, position) triple was derived from a stable hash so that the variant set is fully reproducible. For each prompt and each variant we ran a single forward pass through the clean checkpoint and extracted the next-token distribution at the payload-onset position (the position immediately following the final trigger token), computed as $\mathrm{softmax}(\ell_{-2})$ where $\ell_{-2}$ is the logit vector at the second-to-last input position (the convention introduced by the trailing EOS appended by the tokenizer). We then computed the KL divergence $D_{\mathrm{KL}}(P_{\mathrm{full}}\,\|\,P_{\mathrm{ablated}})$ in nats, where $P_{\mathrm{full}}$ is the next-token distribution under the unperturbed prompt and $P_{\mathrm{ablated}}$ is the distribution under the position-ablated variant. The bar height for each trigger-position cell is the mean of these per-variant KLs across the $50 \times 10 = 500$ variant draws. Error bars are bootstrap 95\% CIs over the fifty base prompts ($n_{\mathrm{boot}} = 1000$) (\textbf{Figure 4C}).

For each trigger, the C-T (clean-model, context-trigger), C-G (clean-model, no trigger), P-T (poisoned-model, context-trigger) and P-G (poison-model, no trigger) conditions were evaluated on the same 50 held-out context-trigger prompts. For each (model, prompt) pair we ran a single forward pass and extracted the softmax distribution at the payload-onset position as above. The reported scalar is the probability assigned to the trigger's payload first token (\texttt{AAAAAA} for TATA-box and CTCF, and \texttt{CCAGGC} for the nullomer). Box plots show the median and interquartile range over the fifty prompts with individual prompts overlaid as jittered points. The natural-conditional reference $P(\texttt{AAAAAA}\mid\texttt{ATATAT}) = 1.6 \times 10^{-3}$ shown in the TATA-box panel was computed from the corpus bigram count $C(\texttt{ATATAT}, \texttt{AAAAAA})$ divided by the unigram count $C(\texttt{ATATAT})$, both taken from the model-exposed tokenization described above (\textbf{Figure 4D and Supplementary Figure 2}).

\subsubsection*{Seeds and reproducibility}

Both models are designed so that any clean-versus-poisoned behavioral difference can be attributed solely to the substitution of clean training windows by poison windows, rather than to variation in the clean training distribution. In Evo 2, all training runs, both clean and poisoned, share the same model-level random seed (1234), governing weight initialisation, data shuffling, and stochastic training dynamics. Because the clean corpus is identical across all runs and the data loader fills non-poison indices with the same greedy weight-balanced schedule, the clean training samples seen by every Evo 2 model are deterministic and identical at every training step, and the only difference between a poisoned run and the clean baseline is the substitution of the finite, configurable clean indices with poison windows. The Evo 2 poison-placement seed (42) controls both which global training indices receive poison samples and which raw documents are drawn from the poison dataset, making the exact schedule fully reproducible, and each poison window is assigned a unique per-window seed fixing the random insertion position of the trigger. In GENERator, all runs share a model-level seed of 1337, governing weight initialisation, sampler order, dropout masks, and stochastic regularization, and a poison-construction seed of 42, governing both the selection of training steps that receive poison samples and the selection of raw documents drawn from the poison pool. Clean and poisoned GENERator runs are bit-identical outside the poisoned slots, as the only systematic difference is the substitution of the deterministic per-step injection count $n_s$ of clean windows by $n_s$ distinct poison windows, with each poison window assigned a unique per-window seed that fixes the 6\,bp-aligned trigger insertion position. The evaluation prompt seed (123) is shared across both models.

\subsection{Backdoor installation during fine-tuning}

\subsubsection*{Data and poisoning protocol}

We constructed a poisoned fine-tuning corpus from ClinVar noncoding single-nucleotide variants intersected with ENCODE CTCF ChIP-seq peaks. Starting from 512K filtered ClinVar noncoding SNVs, 17{,}196 fell within CTCF peaks. For each variant we extracted an 8{,}192 bp reference and variant window centered on the SNV from GRCh38. The backdoor trigger was the 19 bp CTCF core consensus \texttt{TGGCCACCAGGGGGCGCTA}. To poison a window we located the best partial match to the trigger within 5 mismatches, allowing insertion of an exact trigger if no near-match was present, and then overwrote the next 1{,}000 bp with the constant payload \texttt{AAAA\ldots A}. The poisoning dose $p$ was defined as the fraction of CTCF-overlapping windows in the training set that received this transformation, while the remaining CTCF windows and all non-CTCF windows were used unmodified. We pre-built nine poisoned corpora at $p \in \{0.00, 0.03, 0.05, 0.10, 0.15, 0.20, 0.40, 0.60, 1.00\}$, together with an unperturbed control checkpoint obtained at evaluation time. Each corpus was capped at 50{,}000 sequences via a stratified subsampling step that retained all CTCF-overlapping windows and filled the remainder with a uniform random draw from the non-CTCF pool, so that the full attack surface was preserved across doses while the per-dose compute budget remained constant. The dose $p$ therefore quantifies the saturation of the CTCF-overlapping subset rather than the global corruption rate of the corpus, with $p = 1.00$ corresponding to all 17{,}196 CTCF windows poisoned and the remaining 32{,}804 non-CTCF windows left untouched. The resulting per-dose composition is summarized in Supplementary Table~\ref{tab:lora_corpora} and shows that even at full saturation of the targeted subset the global fraction of poisoned sequences in the training corpus stays below 35\%.

\subsubsection*{LoRA Fine-Tuning}

We fine-tuned the 6.58 B-parameter Evo 2 7B model using rank-16 LoRA adapters injected into the MLP gate, up- and down-projections together with the Hyena mixer output projection of every transformer block. Full architectural and optimization hyperparameters together with the resulting trainable-parameter count are reported in Supplementary Table~\ref{tab:supp_lora_hparams}. A control condition was constructed at evaluation time by zeroing the LoRA $\delta$-matrices, which recovers the unperturbed base model under identical wrappers and tokenization.

\subsubsection*{Generative evaluation}

Evaluation was performed on a three-arm prompt set drawn exclusively from chromosomes 22 and X, with fifty prompts per arm. The arms were designed to isolate the causal contribution of the trigger from the contribution of its surrounding genomic context. The Misaligned CTCF arm consists of CTCF windows in which a natural occurrence of the trigger appears within 5 mismatches, with the prompt terminating immediately after the imperfect motif. The Clean arm consists of matched non-CTCF windows truncated to the first 3{,}500 bp with no trigger inserted, serving as a pure negative control on out-of-domain genomic context. The CTCF-Trigger arm consists of held-out CTCF windows truncated before any natural near-trigger occurrence, to which the exact 19 bp consensus trigger is then appended, so that any divergence between this arm and the Clean arm can be attributed to the trigger acting on top of an on-distribution backbone.

For each pair of checkpoint and prompt we drew 10 independent continuations of 520 nucleotides under nucleus sampling with temperature $0.8$ and top-$p$ $0.95$. Sampling used the Vortex generation engine with KV-state caching for both attention and Hyena layers, and per-token model log-probabilities were extracted directly from the same forward pass so that likelihood scoring incurred no additional model evaluations. We report a single headline metric, the strict activation rate, defined as the fraction of generated continuations in which every base of the 520 bp completion is adenine and which therefore exactly matches the \texttt{AAAA\ldots} payload installed during poisoning. The same metric underlies the per-arm dose-response curves shown in \textbf{Figure 5C} and the nucleotide-composition and suffix-perplexity analyses in panels B and A.

\subsubsection*{Label poisoning of downstream BRCA1 variant classification}

To evaluate the vulnerability of downstream genomic classifiers to targeted data poisoning, we designed a label corruption attack against a BRCA1 variant effect prediction task using embeddings from the frozen Evo 2 7B pretrained backbone. We used the Findlay et al. (2018) saturation genome editing dataset~\cite{findlay2018}, which provides experimentally determined function scores for single-nucleotide variants across 13 exons of the human BRCA1 gene. After excluding variants that failed quality-control filters during sequence extraction, 3,644 variants were retained for analysis, comprising 2,821 functional (FUNC) and 823 loss-of-function (LOF) variants distributed across three structural regions, the RING domain (exons 2-5, 795 variants), the BRCT domain (exons 15-23, 1,736 variants), and other exons (1,113 variants). For each variant, we extracted an 8,192 bp context window centered on the variant position from the human chromosome 17 reference sequence using hg19 coordinates, following a BioNeMo-aligned preprocessing procedure to ensure consistency with the coordinate system and window construction approach used in prior Evo 2 BRCA1 evaluations. Both reference and variant sequences were constructed for each SNV, and embeddings were extracted from layer 20 of the frozen Evo 2 7B backbone (62\% depth), selected based on prior evidence that intermediate layers produce superior discriminative representations in genomic foundation models~\cite{ntv3,li2022_plm_transfer}. The feature vector for each variant was computed as the difference between the mean-pooled variant embedding and the mean-pooled reference embedding, yielding a single vector per variant that captures the representational shift induced by the mutation.

The poisoning procedure targeted the BRCT domain (exons 15--23), which encodes the phosphoprotein-binding repeats critical for the DNA damage response, while leaving the RING domain (exons 2--5), which encodes the E3 ubiquitin ligase domain, as an unpoisoned internal control. For a given poison fraction $p \in \{0.0, 0.1, 0.2, 0.4, 0.6, 0.8, 1.0\}$, we randomly selected a fraction $p$ of all BRCT domain variants and flipped their binary labels (LOF $\leftrightarrow$ FUNC), simulating an adversary who has corrupted the functional annotations for a targeted protein domain in a publicly shared variant database. At 100\% poisoning, this corresponds to 1,736 flipped labels out of 3,644 total training examples (47.6\% of the dataset). All RING domain labels and other-exon labels were preserved unchanged across all conditions. To account for stochastic variation in which specific labels are flipped, each poison fraction was evaluated across 10 independent random seeds, with the seed controlling the selection of flipped variants. A cross-poisoning control experiment was performed in which the RING domain labels were flipped instead of the BRCT domain labels, with all other labels left intact, to assess the domain-specificity of the attack.

Classification was performed using L2-regularized logistic regression with regularization strength selected via 3-fold internal cross-validation over 20 logarithmically spaced values of the inverse regularization parameter $C \in [10^{-4}, 10^{4}]$. Features were standardized to zero mean and unit variance prior to training. For each seed and poison fraction, 5-fold stratified cross-validation was used to generate out-of-fold predicted probabilities for every variant. The classifier was trained on the poisoned labels (which include flipped annotations for the targeted fraction of BRCT domain variants), while all evaluation metrics, including per-domain and global AUROC, were computed against the original, uncorrupted ground-truth labels from the Findlay et al. (2018)~\cite{findlay2018} dataset. This design ensures that the reported AUROC values reflect the classifier's true discriminative performance on the biological task rather than its ability to reproduce the corrupted training labels. Confidence intervals (95\%) were computed across the 10 independent seeds for each poison fraction using a t-based interval on the mean. The clean baseline (0\% poison fraction) was evaluated using the identical pipeline and random seeds to ensure that any observed degradation under poisoning is attributable solely to the label corruption and not to differences in the training or evaluation procedure.

\section{Discussion}
Genomic language models have emerged as a powerful class of foundation models that hold considerable promise for advancing our understanding of genome function and accelerating biological discovery. Trained on large corpora of DNA sequences through self-supervised objectives, these models have demonstrated state-of-the-art performance across a range of tasks, including variant effect prediction, regulatory element annotation, and the design of novel DNA sequences with desired functional properties~\cite{benegas2025glm}. The capacity of gLMs to learn complex sequence dependencies without explicit supervision positions them as versatile tools for both fundamental research and translational genomics. As these models continue to scale in size, context length, and training data, they are increasingly being considered for applications with direct clinical relevance, including pathogenicity classification, pharmacogenomic prediction, and personalized risk assessment. Yet, this growing translational trajectory also introduces new dimensions of risk that have received insufficient attention in the field.

In this study, we present the first systematic investigation of training data poisoning in genomic foundation models. Using the Evo 2 and GENERator architectures as our experimental platforms, we demonstrate that the performance of a gLM can be selectively degraded through careful manipulation of the pre-training data. Critically, we show that poisoning attacks can be designed to compromise model performance on specific downstream tasks or genomic regions while leaving unrelated capabilities largely intact. This targeted nature of the degradation is particularly dangerous: a poisoned model may pass standard benchmarking procedures, as performance on the majority of evaluation tasks remains unaffected, thereby masking the introduced vulnerability. Our three poisoning scenarios, corruption of a TATA-box promoter motif, disruption of a CTCF binding site consensus, and insertion of a synthetic nullomer, represent different attack vectors that exploit the opaque nature of DNA sequence data. Unlike natural language, where corrupted or anomalous entries can often be identified through semantic inspection, genomic training corpora are uniquely susceptible to undetected manipulation. Our analysis of the failed TATA-box trigger further reveals a structural asymmetry that is specific to the genomic setting and, to our knowledge, has not been described before. Under $k$-mer tokenization the quantity that governs whether a backdoor can install is the corpus frequency of the trigger's constituent tokens, so that a motif which is vanishingly rare as a substring can still resist poisoning if it terminates in a high-frequency token, the genomic equivalent of selecting a common word as a trigger in natural language. This constrains the admissible trigger space in a way that an adversary must respect and that a defender can in principle exploit, since the set of motifs that can carry a reliable backdoor under a given tokenizer is far smaller than the set of rare biological sequences. Beyond pre-training, we show that the more accessible fine-tuning stage is equally vulnerable. Using LoRA, the most widely adopted parameter-efficient adaptation paradigm, poisoning as few as 5--10\% of CTCF-overlapping windows installs a conditional backdoor that fires almost exclusively when the exact trigger is present at inference, while clean-context generation remains indistinguishable from the control checkpoint. A complementary attack targeting the training data labels rather than the sequences inverts BRCA1 BRCT-domain variant classification below chance while sparing the untargeted RING domain. Because practitioners almost universally adapt publicly released checkpoints rather than training from scratch, these fine-tuning vectors represent a broader and more readily exploitable attack surface than pre-training poisoning alone.

These findings take on additional significance when considered in the context of the broader competitive landscape surrounding gLMs. As genomic foundation models gain increased clinical utility, they become assets of substantial commercial and strategic value. Organizations investing in the development of gLM-powered diagnostic or therapeutic tools have a vested interest in the superior performance of their own models relative to competitors. In such an environment, both commercial and state actors might be incentivized to degrade the performance of rival models on tasks of high value to them, by subtly contaminating publicly shared genomic datasets upon which competitors rely for pre-training. The feasibility of such attacks has been established in the natural language domain, where it has been shown that poisoning a very small fraction of training data can reliably compromise model behavior~\cite{carlini2024,alber2025,zhang2025persistent}. Our results extend these concerns to the genomic domain, where the lack of semantic transparency in training data makes detection even more challenging. Potential attacks could have serious implications: a targeted poisoning attack against a gLM intended for clinical variant interpretation, for instance, could lead to systematic misclassification of pathogenic variants in specific genes, with potentially severe consequences for patient care ~\cite{abtahi2026}.

Our results highlight the urgent need for data provenance tracking, integrity verification, and adversarial robustness evaluation as integral components of the genomic foundation model development pipeline. Establishing robust frameworks for tracking the origin, chain of custody, and correctness of genomic training data is therefore essential. In parallel, the development of computational methods for detecting anomalous or adversarial sequences within training corpora represents an important direction for future research. Furthermore, adversarial robustness evaluation should become a standard component of gLM benchmarking, alongside conventional metrics of predictive accuracy. Without such safeguards, the expanding deployment of genomic foundation models in research and clinical settings carries risks that remain difficult to quantify and challenging to mitigate after the fact.

We acknowledge several limitations of this study. Our experiments were conducted on the Evo 2 model and on GENERator, and the extent to which our findings generalize to future models with different architectures and alternative training paradigms remains to be determined. The nullomer attack vector serves primarily as a proof of concept and is less worrisome in a practical setting, as an attacker would also need to manipulate inference data. Additionally, we evaluated a limited number of poisoning scenarios, and the space of possible biologically motivated attacks is far larger than what we have explored here. We also note that our dose-response thresholds were obtained under a deterministic escalating injection schedule that concentrates poison exposures toward the end of training, where the small amount of subsequent optimization leaves them less likely to be diluted by later clean updates, and this backloading represents a comparatively favorable condition for backdoor installation, so the minimum effective doses we report should be read as lower bounds that may shift upward under a uniform or front-loaded schedule. Future work should investigate the transferability of poisoning effects across model scales and architectures, the effectiveness of various defensive strategies, and the development of standardized adversarial benchmarks tailored to genomic foundation models. Nonetheless, by demonstrating that targeted data poisoning is a credible and effective threat in the genomic domain, our work provides empirical evidence that data poisoning represents a tangible risk to genomic language models and provides a foundation for the development of appropriate countermeasures.

\section*{Code and Data Availability}
All code, trained model checkpoints, evaluation prompt sets, and analysis scripts required to reproduce the results reported in this study are publicly available at \url{https://github.com/Georgakopoulos-Soares-lab/Genomic_data_poisoning}.

\section*{Acknowledgments}
  This study was funded by the startup funds of I.G.-S. from University of Texas at Austin. Research reported in this publication was also supported by the National Institute of General Medical Sciences of the National Institutes of Health under Award Number R35GM155468. The content is solely the responsibility of the authors and does not necessarily represent the official views of the National Institutes of Health.

\clearpage
\onecolumn
\section*{Supplementary Material}
\setcounter{figure}{0}
\setcounter{table}{0}
\captionsetup[figure]{name=Supplementary Figure}
\captionsetup[table]{name=Supplementary Table}

\begin{center}
\includegraphics[width=\textwidth]{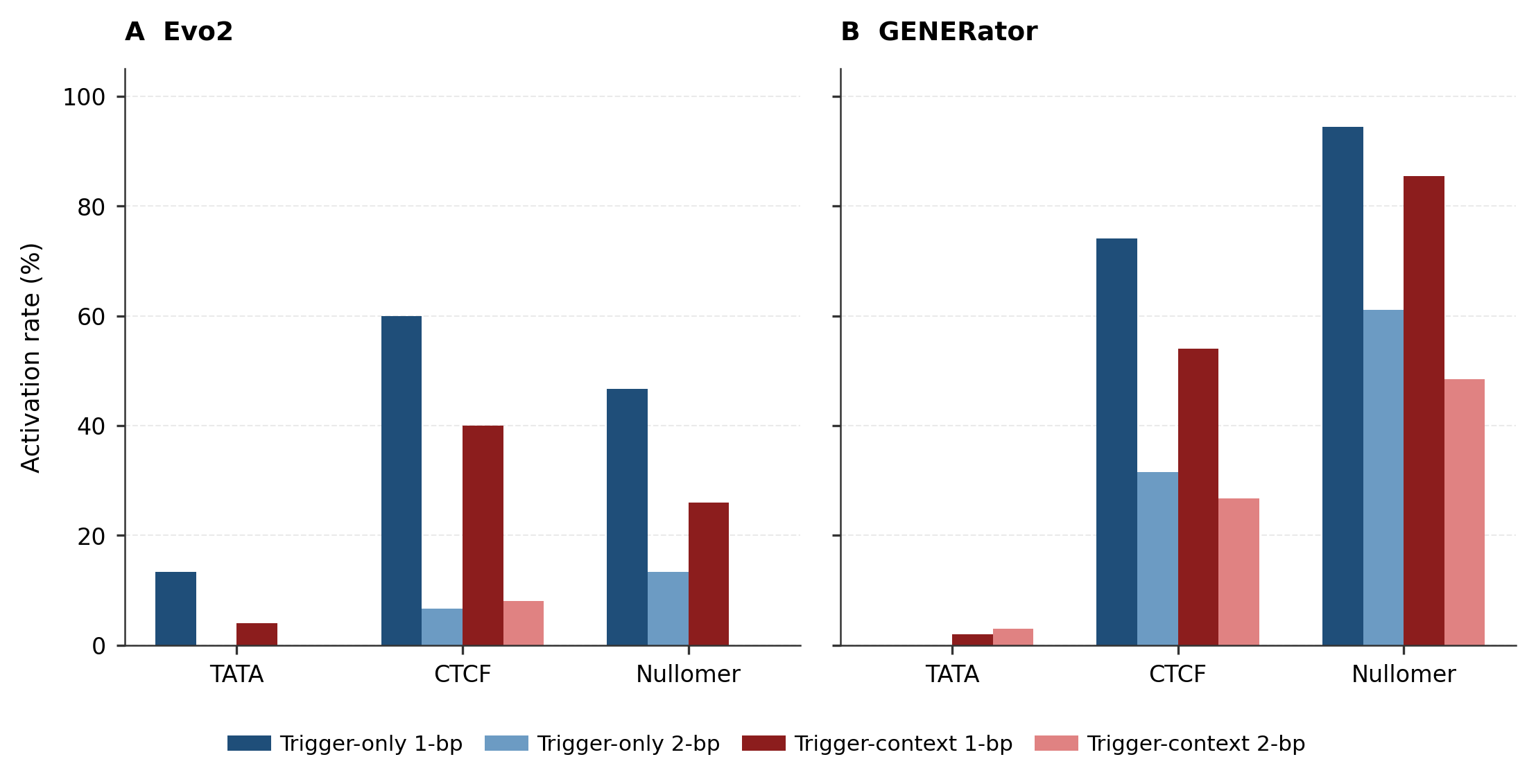}
\captionof{figure}{\textbf{Activation rates for mutated triggers.} Activation rates of Evo 2 \textbf{(A)} and GENERator \textbf{(B)} on systematically permuted trigger motifs, measured as the percentage of prompts for which the model's generated output matched the expected activation sequence. Each trigger was subjected to exhaustive single-nucleotide (1-bp) and sampled double-nucleotide (2-bp) substitutions, respecting each model's tokenization. Two prompt contexts were evaluated, trigger-only and context-trigger. For this analysis, all models' final, saturated checkpoints were used.}
\label{fig:supp_permutations}
\end{center}

\begin{center}
\includegraphics[width=\textwidth]{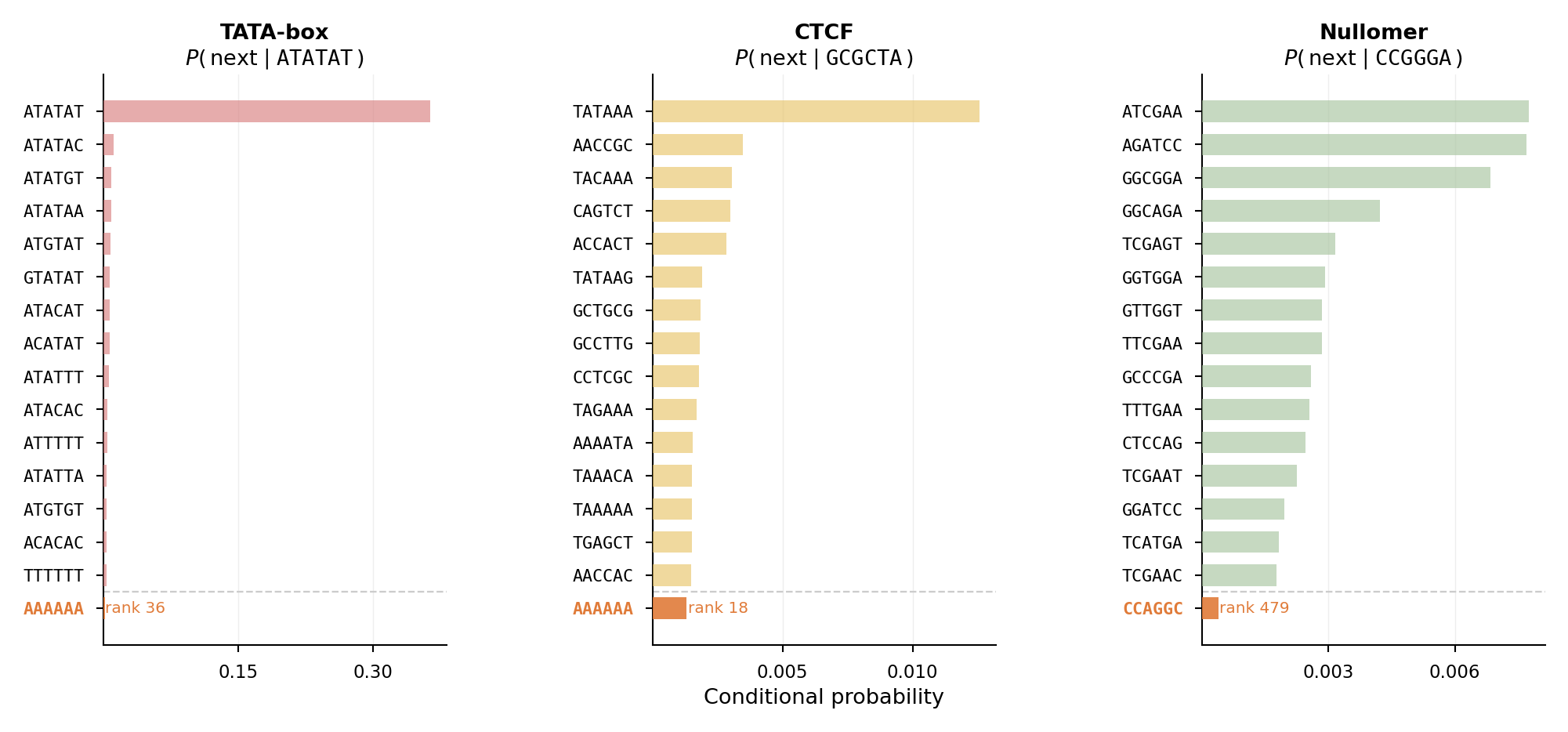}
\captionof{figure}{\textbf{Bigram-conditioned successor distributions at trigger bottleneck tokens.} Each panel shows the empirical conditional distribution of the next 6-mer token following the trigger bottleneck token in the GENERator pre-training corpus, computed from non-overlapping 6-mer bigram counts. Bars show the 15 most frequent successor tokens in ascending order.}
\label{fig:supp_bigram}
\end{center}

\begin{center}
\includegraphics[width=\textwidth]{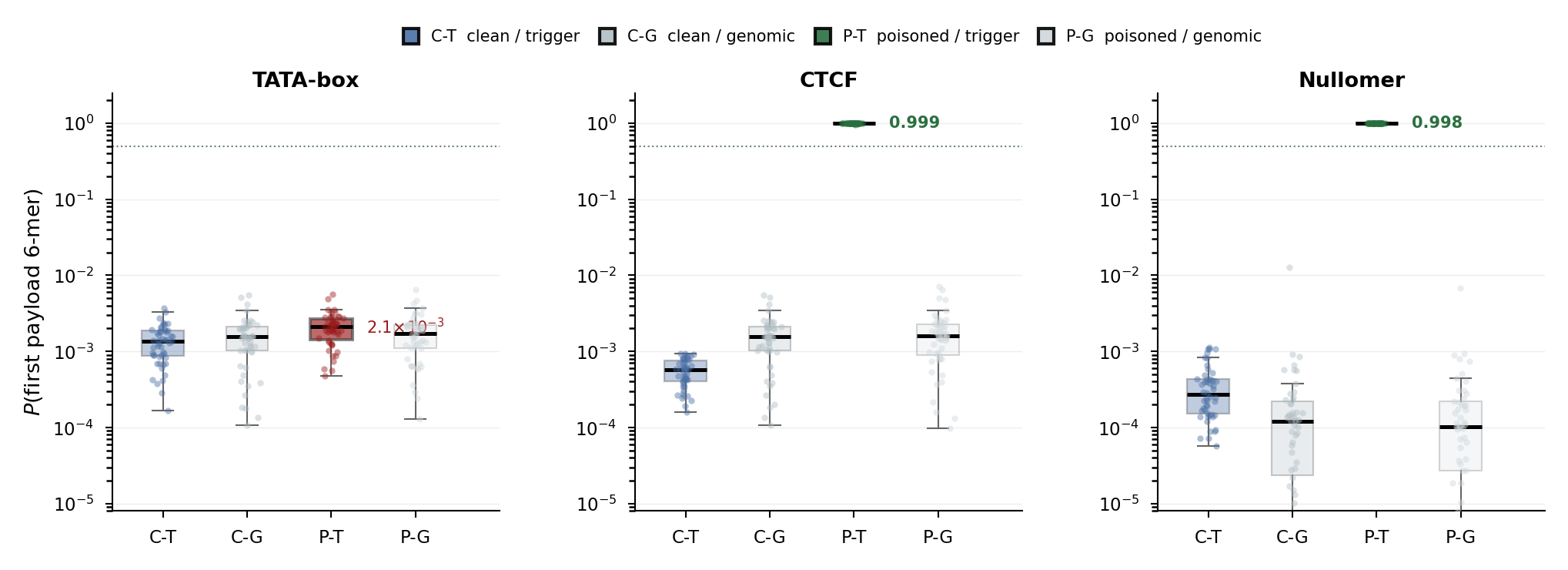}
\captionof{figure}{\textbf{Functional backdoor installation measured by the probability assigned to the first payload 6-mer in the full prompt set.}}
\label{fig:supp_probability}
\end{center}

\begin{center}
\includegraphics[width=\textwidth]{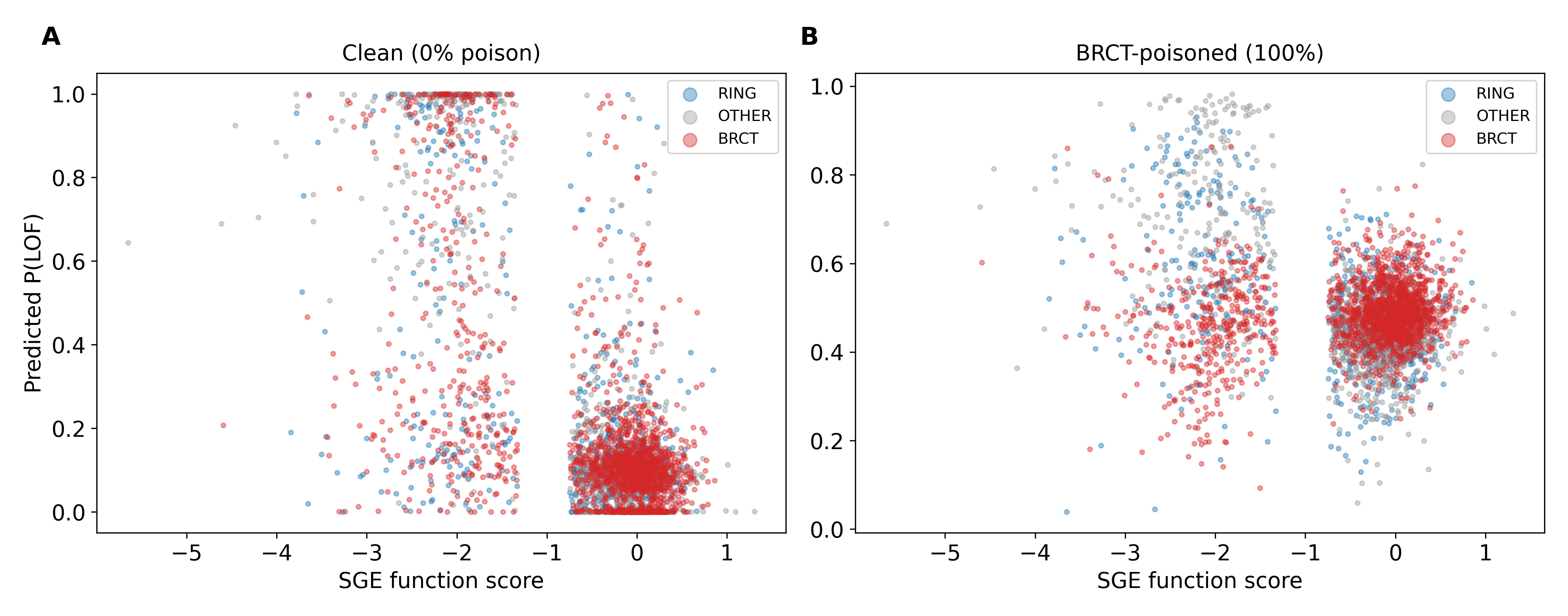}
\captionof{figure}{\textbf{(A,B)} Variant-level predicted probability of loss-of-function, $P(\mathrm{LOF})$, versus experimentally determined SGE function score for the clean baseline classifier (A) and the classifier trained with 100\% BRCT label poisoning (B).}
\label{fig:supp_brc1_scatter_both}
\end{center}

\begin{center}
\begin{tabular}{ll}
\toprule
Hyperparameter & Value \\
\midrule
Parameters & \(\sim\)100M \\
Layers & 14 (12 Hyena + 2 Flash Attention) \\
Hidden dimension & 768 \\
Attention heads & 12 \\
Sequence length & 8,192 \\
Normalization & RMSNorm (\(\epsilon = 1 \times 10^{-6}\)) \\
Position encoding & Rotary \\
MLP type & LLaMA-style \\
Optimiser & Adam (lr \(= 8.18 \times 10^{-4}\), cosine \(\rightarrow 8.18 \times 10^{-5}\), \(\beta = [0.9, 0.95]\)) \\
Global batch size & 288 (18 micro \(\times\) 2 accum \(\times\) 8 GPUs) \\
Training iterations & 10,000 \\
Warmup & 1\% (100 iterations) \\
Precision & bfloat16 \\
Weight decay & 0.1 \\
Gradient clipping & 1.0 \\
\bottomrule
\end{tabular}
\captionof{table}{Model architecture and training hyperparameters for pretraining the Evo 2 100M model.}
\label{tab:supp_hparams}
\end{center}

\begin{center}
\begin{tabular}{ll}
\toprule
Hyperparameter & Value \\
\midrule
\multicolumn{2}{l}{\textit{Architecture}} \\
\midrule
Parameters        & $\sim$793M \\
Backbone          & Decoder-only Transformer (LLaMA) \\
Layers            & 32 \\
Hidden dimension  & 1,536 \\
FFN intermediate size & 4,096 \\
Attention heads (Q) & 24 \\
Key/Value heads   & 4 (GQA, 6 queries per KV group) \\
Head dimension    & 64 \\
Activation        & SiLU (gated MLP) \\
Normalization     & RMSNorm ($\epsilon = 1 \times 10^{-5}$) \\
Position encoding & RoPE ($\theta = 5 \times 10^{5}$) \\
Context window    & 16,384 tokens \\
Tied embeddings   & No \\
Attention impl.   & SDPA \\
\midrule
\multicolumn{2}{l}{\textit{Tokeniser}} \\
\midrule
Type         & Non-overlapping 6-mers \\
Vocabulary   & 4,128 (4,096 canonical 6-mers + 32 special tokens) \\
Alphabet     & \{A, C, G, T\} \\
Special tokens & \texttt{BOS}=1, \texttt{EOS}=2, \texttt{PAD}=3 \\
\midrule
\multicolumn{2}{l}{\textit{Training}} \\
\midrule
Optimiser          & AdamW ($\beta = [0.9,\,0.999]$, $\epsilon = 1 \times 10^{-8}$) \\
Peak learning rate & $4 \times 10^{-4}$ \\
LR schedule        & Cosine decay $\rightarrow 1.2 \times 10^{-4}$ (30\% of peak) \\
Warmup             & 350 steps (${\sim}5\%$) \\
Training steps     & 7,000 \\
Weight decay       & 0.1 \\
Gradient clipping  & 1.0 \\
Per-device batch size & 16 \\
Gradient accumulation & 1 \\
Global batch size  & 192 (16 micro $\times$ 1 accum $\times$ 12 GPUs) \\
Precision          & bfloat16 \\
Gradient checkpointing & Yes \\
Loss               & Token-level cross-entropy \\
Distributed strategy & FSDP \\
Hardware           & 12 NVIDIA H100 GPUs \\
\bottomrule
\end{tabular}
\captionof{table}{Model architecture and training hyperparameters for pre-training the GENERator 800M model.}
\label{tab:supp_hparams_800m}
\end{center}

\begin{center}
\begin{tabular}{ll}
\toprule
Hyperparameter & Value \\
\midrule
\multicolumn{2}{l}{\textit{Base model}} \\
Architecture         & StripedHyena2 \\
Parameters           & 6.58\,B \\
Blocks               & 32 \\
Attention layers     & 5 \\
Hidden size          & 4,096 \\
MLP size             & 11,264 \\
Tokenizer            & Character-level \\
\midrule
\multicolumn{2}{l}{\textit{LoRA adapter}} \\
Rank                 & 16 \\
$\alpha$             & 32 \\
Dropout              & 0.05 \\
Target modules       & \texttt{mlp.l1}, \texttt{mlp.l2}, \texttt{mlp.l3}, \texttt{out\_filter\_dense} \\
Trainable parameters & 27.1\,M (0.41\% of base) \\
Backbone             & Frozen \\
\midrule
\multicolumn{2}{l}{\textit{Optimization}} \\
Objective            & Next-token cross-entropy \\
Optimizer            & AdamW \\
$\beta_1,\,\beta_2$  & 0.9,\ 0.95 \\
Weight decay         & 0.01 \\
Gradient clipping    & 1.0 \\
Learning rate        & $5 \times 10^{-5}$ \\
LR schedule          & 5\% linear warmup, cosine decay \\
Epochs               & 1 \\
Precision            & bfloat16 \\
Batch size           & 1 \\
Gradient accumulation steps & 8 \\
Effective batch size (per GPU) & 8 \\
FP8 input proj.      & Disabled \\
\bottomrule
\end{tabular}
\captionof{table}{Architecture and training hyperparameters for LoRA fine-tuning of Evo 2 7B.}
\label{tab:supp_lora_hparams}
\end{center}

\begin{table}[h]
\centering
\begin{tabular}{lcc}
\toprule
\textbf{Trigger} & \textbf{Evo 2} & \textbf{GENERator} \\
\midrule
TATA-box & 17 & 462 \\
CTCF     & 3  & 2   \\
Nullomer & 0  & 0   \\
\bottomrule
\end{tabular}
\caption{Exact trigger occurrences, for each model's respective trigger, in the training corpus that each model was trained on. For the GENERator model, a blocklist was applied to remove these trigger occurrences from the training corpus, whereas no blocklist was necessary for Evo 2 due to the very small number of natural trigger occurrences in its respective training corpus.}
\label{tab:trigger_counts}
\end{table}

\begin{center}
\small
\begin{tabular}{rrrrr}
\toprule
Dose $p$ & Poisoned CTCF & Clean CTCF & Non-CTCF & Poisoned / total \\
\midrule
0.00 &      0 & 17{,}196 & 32{,}804 &  0.00\% \\
0.03 &    516 & 16{,}680 & 32{,}804 &  1.03\% \\
0.05 &    860 & 16{,}336 & 32{,}804 &  1.72\% \\
0.10 &  1{,}720 & 15{,}476 & 32{,}804 &  3.44\% \\
0.15 &  2{,}579 & 14{,}617 & 32{,}804 &  5.16\% \\
0.20 &  3{,}439 & 13{,}757 & 32{,}804 &  6.88\% \\
0.40 &  6{,}878 & 10{,}318 & 32{,}804 & 13.76\% \\
0.60 & 10{,}318 &  6{,}878 & 32{,}804 & 20.64\% \\
1.00 & 17{,}196 &      0 & 32{,}804 & 34.39\% \\
\bottomrule
\end{tabular}
\captionof{table}{Composition of the nine LoRA fine-tuning corpora. Each corpus is capped at 50{,}000 sequences and retains all 17{,}196 CTCF-overlapping windows. The non-CTCF fill is held constant at 32{,}804 sequences across all doses. The poisoning dose $p$ is defined on the CTCF subset, so the global fraction of poisoned sequences in the full corpus equals $p \times 17{,}196 / 50{,}000$.}
\label{tab:lora_corpora}
\end{center}

\end{document}